\documentclass[aps,superscriptaddress,nofootinbib,11pt]{revtex4-1} 
\pdfoutput=1

\usepackage{dcolumn}    
\usepackage{pstricks}
\usepackage{color}
\usepackage{graphicx}
\usepackage{amsmath}
\usepackage{mathtools}
\usepackage[T1]{fontenc}
\usepackage[utf8]{inputenc}
\usepackage{ae}
\usepackage{amssymb}
\usepackage[colorlinks=true,allcolors=darkred, pdfborder={0 0 0}]{hyperref}

\usepackage{booktabs}

\allowdisplaybreaks

\def\MGUT{M_\mathrm{GUT}}
\def\SO10{\mathrm{SO}(10)}

\newcommand{\rep}[1]{\mathbf{#1}}
\newcommand{\repb}[1]{\mathbf{\overline{#1}}}

\definecolor{darkred}{rgb}{0.6,0,0}

\begin{document}

\title{Fits to Non-Supersymmetric SO(10) Models with Type I and II Seesaw Mechanisms Using Renormalization Group Evolution}	
	
\author{Tommy Ohlsson}
\email{tohlsson@kth.se}
\affiliation{Department of Physics,
	School of Engineering Sciences,
	KTH Royal Institute of Technology,
	AlbaNova University Center,
	Roslagstullsbacken 21,
	SE--106 91 Stockholm,
	Sweden}
\affiliation{The Oskar Klein Centre for Cosmoparticle Physics,
	AlbaNova University Center,
	Roslagstullsbacken 21,
	SE--106 91 Stockholm,
	Sweden}
\affiliation{University of Iceland, 
	Science Institute, 
	Dunhaga 3, 
	IS--107 Reykjavik, 
	Iceland}

\author{Marcus Pernow}
\email{pernow@kth.se}
\affiliation{Department of Physics,
	School of Engineering Sciences,
	KTH Royal Institute of Technology,
	AlbaNova University Center,
	Roslagstullsbacken 21,
	SE--106 91 Stockholm,
	Sweden}
\affiliation{The Oskar Klein Centre for Cosmoparticle Physics,
	AlbaNova University Center,
	Roslagstullsbacken 21,
	SE--106 91 Stockholm,
	Sweden}

\begin{abstract}
We consider numerical fits to non-supersymmetric $\SO10$-based models in which neutrino mass is generated by the type-I or type-II seesaw mechanism or a combination of both. The fits are performed with a sophisticated top-down procedure, taking into account the renormalization group equations of the gauge and Yukawa couplings, integrating out relevant degrees of freedom at their corresponding mass scales, and using recent data for the Standard Model observables. We find acceptable fits for normal neutrino mass ordering only and with neutrino mass generated by either type-I seesaw only or a combination of types I and II seesaw in which type-I seesaw is dominant. Furthermore, we find predictions from the best fit regarding the small neutrino masses, the effective neutrinoless double beta decay mass, and the leptonic CP-violating phase. Finally, we show that the fits are rather insensitive to the chosen value of the unification scale.
\end{abstract}

\maketitle

\section{Introduction}

Grand unified theories (GUTs)~\cite{Georgi:1974sy}, and in particular their non-supersymmetric (non-SUSY) $\SO10$ realizations~\cite{Fritzsch:1974nn}, embed of the Standard Model (SM) group in a unifying gauge group. Thereby, they lead to frameworks for physics beyond the SM which may address some of the outstanding problems, both phenomenological and aesthetic. In particular, $\SO10$ models account for the generation of neutrino mass in a natural way through the type-I~\cite{Minkowski:1977sc,GellMann:1980vs,Mohapatra:1979ia,Yanagida:1979,Schechter:1980gr} and type-II~\cite{Magg:1980ut,Lazarides:1980nt,Mohapatra:1980yp} seesaw mechanisms. To verify the viability of these models, one must attempt to fit their parameters to the known observables of the SM in order to find whether or not the models in question allow the observed low-energy values.

Fits of the Yukawa sector of $\SO10$ models to the observables of the SM have previously been presented in the literature, with various levels of detail. As neutrino data became available to the level of precision that allowed fits to be made, the initial attempts to accommodate neutrino masses and leptonic mixing parameters were performed for SUSY models~\cite{Bertolini:2004eq,Babu:2005ia,Bertolini:2006pe,Bajc:2008dc,Altarelli:2010at,Fukuyama:2015kra} with the type-I or type-II seesaw mechanism, or a mixture thereof. There have also been numerous fits of the Yukawa sector of various non-SUSY models with type-I seesaw~\cite{Altarelli:2013aqa,Dueck:2013gca,Meloni:2014rga,Babu:2015bna,Meloni:2016rnt,Ohlsson:2018qpt,Boucenna:2018wjc}, with the conclusion that these fits are possible depending on the specifics of the Yukawa sector and the symmetry breaking pattern. Furthermore, fits with type-II seesaw or a combination of types I and II seesaw have been considered in non-SUSY scenarios~\cite{Joshipura:2011nn,Babu:2016bmy}, which concluded that models with type-II seesaw only do not yield acceptable fits, but that type-II seesaw in combination with type-I seesaw provides good fits. 

The works mentioned above can be classified based on the level of sophistication of the procedure used to perform the fits. Most of the previous fits have been performed with a \emph{bottom-up} approach, by first evolving the experimental values of the SM observables up to the GUT scale $\MGUT$ using the renormalization group equations (RGEs) and then fitting the $\SO10$ Yukawa sector to the evolved data at that scale. This procedure involves several approximations, since the renormalization group (RG) evolution in general depends on parameters which are not known a priori, such as the mass scales of right-handed neutrinos (where they are integrated out) and the matching conditions at intermediate breaking steps. The more complete procedure is to use a \emph{top-down} approach, which involves randomly sampling the parameters of the $\SO10$ Yukawa sector at $\MGUT$ and evolving each parameter down to the electroweak scale $M_\mathrm{Z}$ using the RGEs, where they are compared to experimental values of the observables, as has been done in Refs.~\cite{Dueck:2013gca,Meloni:2016rnt,Ohlsson:2018qpt,Boucenna:2018wjc}. Related to this is the treatment of the RG evolution of the neutrino sector. As the parameters are evolved down from $\MGUT$ to $M_\mathrm{Z}$, a complete analysis should integrate out the right-handed neutrinos at their respective mass scales, which has been carried out in Refs.~\cite{Dueck:2013gca,Boucenna:2018wjc}. Other works either performed the fits at $\MGUT$ or assumed that all right-handed neutrinos were integrated out simultaneously during the RG evolution.

In this work, we consider fits to non-SUSY $\SO10$ models with neutrino mass being generated by either the type-I or type-II seesaw mechanism or a combination of both, similar to Ref.~\cite{Babu:2016bmy}. The procedure used is similar to that of Ref.~\cite{Dueck:2013gca}, which involves sampling the parameters of the models at $\MGUT$, evolving them down to $M_\mathrm{Z}$ using the RGEs, and comparing the resulting values to data of the observables. The novelty of this work is the combination of the type-I and type-II seesaw mechanisms with a proper and complete treatment of the RG evolution, including integrating out right-handed neutrinos at their respective mass scales. Furthermore, we use updated data for all fermion observables. 

This paper is structured as follows. First, in Sec.~\ref{sec:model}, we present the model that we investigate and the origin of the seesaw mechanism in $\SO10$. Next, in Sec.~\ref{sec:running}, we describe how the parameters of this model are related to those of the SM. Then, in Sec.~\ref{sec:procedure}, we discuss the parametrization and the numerical procedure used. Finally, in Sec.~\ref{sec:results}, we present the results before summarizing our findings and concluding in Sec.~\ref{sec:summary}.

\section{Model}\label{sec:model}

The model that we consider is a non-SUSY $\SO10$ model with each generation of fermions and right-handed neutrinos belonging to a $\rep{16}_F$ representation, whereas the Higgs scalars reside in the $\rep{10}_H$ and $\repb{126}_H$ representations. We also introduce a global $\mathrm{U}(1)_\mathrm{PQ}$ symmetry which has a double purpose. Firstly, it solves the strong CP problem and provides the QCD axions~\cite{Peccei:1977hh,Peccei:1977ur,Weinberg:1977ma,Wilczek:1977pj}. Secondly, and more importantly for the Yukawa sector, it allows us to complexify the real $\rep{10}_H$ representation without introducing additional couplings~\cite{Bajc:2005zf}, as described in Sec.~\ref{sec:SO10}. 

We assume that the $\SO10$ symmetry is broken at $\MGUT=2\times 10^{16}\,\mathrm{GeV}$ in one step to the SM, similarly to Ref.~\cite{Dueck:2013gca}. Since we focus on the fermion observables, the exact details of symmetry breaking and the exact value of $\MGUT$ are irrelevant (and we check this explicitly in Sec.~\ref{sec:results}). Therefore, the gauge couplings in our model do not unify. As in Ref.~\cite{Dueck:2013gca}, we assume that this is taken care of by some new physics between $M_\mathrm{Z}$ and $\MGUT$. For examples of such models, see Refs.~\cite{Frigerio:2009wf,Parida:2016hln,Boucenna:2018wjc}. At $M_\mathrm{Z}$, the electroweak symmetry is further broken according to the usual Higgs mechanism, so that the complete breaking chain is
\begin{equation}
\SO10 \xrightarrow[]{\MGUT}\mathrm{SU}(3)_C\times\mathrm{SU}(2)_L\times\mathrm{U}(1)_Y
\xrightarrow[]{\ \ M_\mathrm{Z}\ \ } \mathrm{SU}(3)_C\times\mathrm{U}(1)_Q.
\end{equation}

\subsection{SO(10) Lagrangian}\label{sec:SO10}
The model described above uniquely defines the Lagrangian of the Yukawa sector to be 
\begin{equation}
\mathcal{L}_Y = \rep{16}_F(Y_{10}\rep{10}_H + Y_{126}\repb{126}_H)\rep{16}_F,
\end{equation}
where $Y_{10}$ and $Y_{126}$ are $3\times3$ matrices in flavor space. The representation $\repb{126}_H$ is complex in $\SO10$, but the representation $\rep{10}_H$ is real. This means that the two $\mathrm{SU}(2)_L$ doublets in the $\rep{10}_H$ necessarily take the same vacuum expectation value (vev), which would imply certain mass relations that contradict data~\cite{Bajc:2005zf}. Thus, we complexify it, which allows the two components to take different vevs. This in effect introduces  a $\rep{10}_H^*$, which may couple to the fermions in the representation $\rep{16}_F$ with a new Yukawa matrix, thereby decreasing the predictability of the model. However, if we assign the global PQ  charges such that 
\begin{equation}
\rep{16}_F\rightarrow\mathrm{e}^{\mathrm{i}\alpha}\rep{16}_F,\quad \rep{10}_H\rightarrow\mathrm{e}^{-2\mathrm{i}\alpha}\rep{10}_H, \quad \repb{126}_H\rightarrow\mathrm{e}^{-2\mathrm{i}\alpha}\repb{126}_H,
\end{equation}
where $\alpha$ is some real parameter, then this additional coupling of the $\rep{10}_H^*$ to the $\rep{16}_F$ is forbidden.

After symmetry breaking, the SM Higgs doublet is a combination of the four $\mathrm{SU}(2)_L$ doublets found in the complexified $\rep{10}_H$ and the $\repb{126}_H$. The Yukawa matrices of the SM are determined by combinations of the Yukawa matrices $Y_{10}$ and $Y_{126}$, weighted by the vevs of these four $\mathrm{SU}(2)_L$ doublets $v^u_{10}$, $v^d_{10}$, $v^u_{126}$, and $v^d_{126}$. Note that these vevs do not exist at the $\SO10$ breaking scale, since they arise from electroweak symmetry breaking, but their values enter as parameters in the matching conditions. Thus, the SM fermion observables may be extracted from the following combinations of $\SO10$ Yukawa matrices $Y_{10}$ and $Y_{126}$~\cite{Altarelli:2013aqa,Dueck:2013gca,Babu:2015bna,Dueck:2013gca,Joshipura:2011nn}
\begin{equation}\label{eq:params}
\begin{split}
Y_u &= \frac{1}{v_\mathrm{SM}}(v^u_{10} Y_{10} + v^u_{126}Y_{126}), \\
Y_d &= \frac{1}{v_\mathrm{SM}}(v^d_{10} Y_{10} + v^d_{126}Y_{126}), \\
Y_\nu &= \frac{1}{v_\mathrm{SM}}(v^u_{10} Y_{10} -3 v^u_{126}Y_{126}), \\
Y_\ell &= \frac{1}{v_\mathrm{SM}}(v^d_{10} Y_{10} -3 v^d_{126}Y_{126}),
\end{split}
\end{equation}
where $Y_u$, $Y_d$, $Y_\nu$, and $Y_\ell$ are the Yukawa matrices for the up-type quarks, the down-type quarks, the neutrinos, and the charged leptons, respectively, while $v_\mathrm{SM}\simeq174\,\mathrm{GeV}$ is the SM Higgs vev. The relative signs and factors of $3$ come from Clebsch--Gordan coefficients. 

\subsection{Seesaw from SO(10)}
To generate neutrino mass via the type-I seesaw mechanism~\cite{Minkowski:1977sc,GellMann:1980vs,Mohapatra:1979ia,Yanagida:1979,Schechter:1980gr}, the right-handed neutrinos must obtain Majorana masses, which occurs naturally in $\SO10$ models~\cite{Lazarides:1980nt}. The right-handed neutrinos reside in the $\rep{16}_F$ and couple to an SM singlet contained in the $\repb{126}_H$. This singlet takes a vev $v^R_{126}$, which generates a Majorana mass matrix
\begin{equation}\label{eq:M_R}
M=v^R_{126}Y_{126}.
\end{equation}
Since $v^R_{126}$ is close to $\MGUT$, the Majorana masses for the right-handed neutrinos are large. Through their Yukawa coupling to the light neutrinos, the type-I seesaw mechanism provides small neutrino masses.

The type-II seesaw mechanism~\cite{Magg:1980ut,Lazarides:1980nt,Mohapatra:1980yp} also naturally occurs in $\SO10$ models~\cite{Bajc:2002iw,Bajc:2004fj} and contributes to neutrino mass. In addition to the SM singlet, the $\repb{126}_H$ contains an $\mathrm{SU}(2)_L$ triplet $\Delta$, which interacts with the light neutrinos with Yukawa coupling
\begin{equation}\label{eq:YDelta}
Y_\Delta = Y_{126}.
\end{equation}
After electroweak symmetry breaking, the interactions between this scalar triplet and the Higgs boson induces a vev $v^L_{126}$ for the triplet. This vev is inversely proportional the square of the scalar triplet mass. Hence, a heavy scalar triplet generates small masses for the neutrinos. 

During RG evolution from $\MGUT$ to $M_\mathrm{Z}$, the right-handed neutrinos and the scalar triplet are integrated out at their respective mass scales. This results in the effective dimension-5 operator for neutrino mass.

\section{SM Observables from SO(10)}\label{sec:running}

In this section, we describe how the sampled parameter values at $\MGUT$ are evolved down to the values of the SM observables at $M_\mathrm{Z}$, as well as the procedure how to integrate out right-handed neutrinos and the scalar triplet at their respective mass scales.

\subsection{Renormalization Group Equations}\label{sec:RGEs}

To perform the RG evolution of the sampled parameters from $\MGUT$ to $M_\mathrm{Z}$, we need to solve the RGEs numerically. The parameters which exhibit RG evolution are the gauge couplings $g_1$, $g_2$, and $g_3$, the Yukawa coupling matrices $Y_u$, $Y_d$, $Y_\nu$, $Y_\ell$, and $Y_\Delta$, the right-handed neutrino Majorana mass matrix $M$, the Higgs quartic coupling $\lambda$, and the coupling matrix of the dimension-5 effective neutrino mass operator $\kappa$ (see Secs.~\ref{sec:typeI} and \ref{sec:typeII}). The general set of RGEs, assuming the presence of the right-handed neutrinos and the scalar triplet, i.e.~with both type-I and type-II seesaw, is~\cite{Jones:1981we,Machacek:1983tz,Machacek:1983fi,Machacek:1984zw,Dueck:2013gca,Antusch:2002rr,Antusch:2005gp,Schmidt:2007nq,Chao:2006ye}
\begin{align}
16\pi^2 \beta_{g_1} &= \frac{41}{10}g_1^3 + \frac{3}{5}g_1^3 = \frac{47}{10} g_1^3,\label{eq:g1}\\
16\pi^2 \beta_{g_2} &= -\frac{19}{6}g_2^3 + \frac{2}{3}g_2^3 = -\frac{5}{2} g_2^3,\label{eq:g2}\\
16\pi^2 \beta_{g_3} &= -7g_3^3, \\
16\pi^2 \beta_{\lambda} &= 6\lambda^2 - 3\lambda \left(3g_2^2 + \frac{3}{5}g_1^2 \right) + 3g_2^4 + \frac{3}{2}\left(\frac{3}{5}g_1^2 + g_2^2 \right)^2 \nonumber \\ & + 4\lambda \mathrm{Tr} \left[ Y_\ell^\dagger Y_\ell + Y_\nu^\dagger Y_\nu + 3Y_d^\dagger Y_d + 3 Y_u^\dagger Y_u \right] \nonumber \\ &- 8\mathrm{Tr}\left[ Y_\ell^\dagger Y_\ell Y_\ell^\dagger Y_\ell + Y_\nu^\dagger Y_\nu Y_\nu^\dagger Y_\nu + 3Y_d^\dagger Y_d Y_d^\dagger Y_d + 3 Y_u^\dagger Y_u Y_u^\dagger Y_u \right], \\
16\pi^2 \beta_{Y_u} &= Y_u \Bigg( \frac{3}{2} Y_u^\dagger Y_u - \frac{3}{2} Y_d^\dagger Y_d - \frac{17}{20} g_1^2 - \frac{9}{4} g_2^2 - 8 g_3^2  +  \mathrm{Tr}\left[ Y_\ell^\dagger Y_\ell + Y_\nu^\dagger Y_\nu + 3Y_d^\dagger Y_d + 3 Y_u^\dagger Y_u\right] \Bigg),\\
16\pi^2 \beta_{Y_d} &= Y_d \Bigg( \frac{3}{2} Y_d^\dagger Y_d - \frac{3}{2} Y_u^\dagger Y_u - \frac{1}{4} g_1^2 - \frac{9}{4} g_2^2 - 8 g_3^2  +  \mathrm{Tr}\left[ Y_\ell^\dagger Y_\ell + Y_\nu^\dagger Y_\nu + 3Y_d^\dagger Y_d + 3 Y_u^\dagger Y_u\right] \Bigg),\\
16\pi^2 \beta_{Y_\nu} &= Y_\nu \Bigg( \frac{3}{2} Y_\nu^\dagger Y_\nu - \frac{3}{2} Y_\ell^\dagger Y_\ell + \frac{3}{2}Y_\Delta^\dagger Y_\Delta - \frac{9}{20} g_1^2 - \frac{9}{4} g_2^2  +  \mathrm{Tr}\left[ Y_\ell^\dagger Y_\ell + Y_\nu^\dagger Y_\nu + 3Y_d^\dagger Y_d + 3 Y_u^\dagger Y_u\right] \Bigg),\\
16\pi^2 \beta_{Y_\ell} &= Y_\ell \Bigg( \frac{3}{2} Y_\ell^\dagger Y_\ell - \frac{3}{2} Y_\nu^\dagger Y_\nu + \frac{3}{2}Y_\Delta^\dagger Y_\Delta - \frac{9}{4} g_1^2 - \frac{9}{4} g_2^2 +  \mathrm{Tr}\left[ Y_\ell^\dagger Y_\ell + Y_\nu^\dagger Y_\nu + 3Y_d^\dagger Y_d + 3 Y_u^\dagger Y_u\right] \Bigg),\\
16\pi^2 \beta_{Y_\Delta} &= \left[ \frac{1}{2} Y_\nu^\dagger Y_\nu + \frac{1}{2} Y_\ell^\dagger Y_\ell + \frac{3}{2} Y_\Delta^\dagger Y_\Delta\right]^T Y_\Delta + Y_\Delta \left[\frac{1}{2} Y_\nu^\dagger Y_\nu + \frac{1}{2} Y_\ell^\dagger Y_\ell + \frac{3}{2} Y_\Delta^\dagger Y_\Delta \right] \nonumber \\ &+ \left[-\frac{3}{2}\left(\frac{3}{5}g_1^2 + 3g_2^2 \right) + \mathrm{Tr}\left( Y_\Delta^\dagger Y_\Delta\right)\right] Y_\Delta, \\
16\pi^2 \beta_{M} &= (Y_\nu Y_\nu^\dagger) M + M(Y_\nu Y_\nu^\dagger)^T, \\
16\pi^2 \beta_{\kappa} &= \frac{1}{2}(Y_\nu^\dagger Y_\nu - 3Y_\ell^\dagger Y_\ell + 3Y_\Delta^\dagger Y_\Delta )^T \kappa + \frac{1}{2}\kappa (Y_\nu^\dagger Y_\nu - 3Y_\ell^\dagger Y_\ell + 3Y_\Delta^\dagger Y_\Delta)  \nonumber \\ &+  2 \mathrm{Tr}\left[Y_\ell^\dagger Y_\ell + Y_\nu^\dagger Y_\nu + 3Y_u^\dagger Y_u + 3Y_d^\dagger Y_d \right]\kappa -3g_2^2\kappa + \lambda\kappa. \label{eq:kappaRGE}
\end{align}

The RGEs in Eqs.~\eqref{eq:g1}--\eqref{eq:kappaRGE} are for the general case in which the set of fields that survive below $\MGUT$ contains both the right-handed neutrinos and the scalar triplet. We are also interested in the situations in which there is no scalar triplet or no right-handed neutrinos below $\MGUT$ (corresponding to pure type-I seesaw or pure type-II seesaw, respectively). In these cases, the equations can be easily modified by removing the irrelevant contributions as follows:
\begin{itemize}
\item {\bf Pure type-I seesaw:} For the case with type-I seesaw only, one removes the second terms from Eqs.~\eqref{eq:g1} and \eqref{eq:g2} as well as any contributions from $Y_\Delta$ in the RGEs. This results in a set of RGEs with no contribution from the scalar triplet $\Delta$. 
\item {\bf Pure type-II seesaw:} For the case with type-II seesaw only, the gauge couplings remain unchanged (since the right-handed neutrinos are neutral under the SM gauge group) and one simply has to remove any contributions from $Y_\nu$ in the RGEs, as well as the RGE for $M$.
\end{itemize} 

\subsection{Right-Handed Neutrino Mass Thresholds}\label{sec:typeI}

For the scenarios in which the right-handed neutrinos are a part of the set of fields below $\MGUT$, the energy scale during the RG evolution of the parameters at some point coincides with the mass of the heaviest right-handed neutrino $N_3$, $\mu=M_3$. At that threshold, we integrate out $N_3$ following the procedure outlined in Refs.~\cite{Antusch:2002rr,Antusch:2005gp}. The procedure entails removing the last row of the matrix $Y_\nu$ (which corresponds to the coupling of the three light neutrinos to $N_3$) and also removing the last row and column of the right-handed neutrino mass matrix $M$. Note that this is basis-dependent, and it is therefore crucial to work in a basis in which the matrix $M$ is diagonal (with the corresponding basis transformation applied to $Y_\nu$). The information regarding the interactions of $N_3$ that is removed from $Y_\nu$ and $M$ is placed in the effective $3\times3$ neutrino mass matrix $\kappa$ as
\begin{equation}
\kappa \rightarrow \kappa + \frac{2}{M_3}\left( Y_\nu^{(3)}\right)^T \left(Y_\nu^{(3)} \right),
\end{equation}
where $Y_\nu^{(3)}$ is the removed row from $Y_\nu$. If this is the first contribution to the effective neutrino mass, $\kappa$ is initially a $3\times 3$ zero matrix. Otherwise, it may be non-zero if there are other contributions to the effective neutrino mass operator that have already been integrated at a higher scale (for example, if we have a scalar triplet $\Delta$ that is heavier than $N_3$). 

Between $M_3$ and the next threshold, we again solve the same equations as above, but with the new matrices $Y_\nu$ and $M$, which are now $2\times 3$ and $2\times 2$ matrices, respectively. Additionally, we have an RGE for $\kappa$ as in Eq.~\eqref{eq:kappaRGE}.

At the threshold of the second right-handed neutrino mass, $\mu=M_2$, we integrate out $N_2$ following the same prescription and updating $\kappa$ as
\begin{equation}
\kappa \rightarrow \kappa + \frac{2}{M_2}\left( Y_\nu^{(2)}\right)^T \left( Y_\nu^{(2)} \right),
\end{equation}
where $Y_\nu^{(2)}$ is the removed row from $Y_\nu$. After this, $Y_\nu$ is a $1\times 3$ matrix and $M$ is a scalar. The same RGEs apply below $M_2$ as above.

At the last right-handed neutrino threshold, $\mu=M_1$, the lightest right-handed neutrino $N_1$ is integrated out and $\kappa$ is updated to 
\begin{equation}
\kappa \rightarrow \kappa + \frac{2}{M_1}\left(Y_\nu^{(1)}\right)^T \left( Y_\nu^{(1)} \right),
\end{equation}
where $Y_\nu^{(1)}$ is the final remaining row from $Y_\nu$. After this threshold, the parameters $Y_\nu$ and $M$ are no longer present in the RG evolution.

\subsection{Triplet Mass Threshold}\label{sec:typeII}

Similarly to the situation described in Sec.~\ref{sec:typeI}, if the scalar triplet $\Delta$ is involved, it has an RGE above its mass threshold. As the RG evolution reaches its mass scale $M_\Delta$, it is integrated out and its interactions with the neutrinos will be encoded in the effective neutrino mass matrix $\kappa$. This contribution takes the form~\cite{Schmidt:2007nq}
\begin{equation}\label{eq:kappaII}
\kappa \rightarrow \kappa -4\frac{v^L_{126}}{v_\mathrm{SM}^2}Y_\Delta.
\end{equation}
Below this mass threshold, $Y_\Delta$ is no longer a parameter of the model and therefore is no longer present in the RG evolution. The RGEs for $g_1$ and $g_2$ are modified accordingly and any contribution from $Y_\Delta$ in the set of RGEs is removed, as noted in Sec.~\ref{sec:RGEs}.

\section{Fitting Procedure}\label{sec:procedure}
In this section, we describe the numerical procedure used, including the parametrization of the $\SO10$ parameters and the input data that we fit to. The input data are based on the experimental values of the SM observables, see Sec.~\ref{sec:ID}.

\subsection{Parametrization}\label{sec:params}

Following the conventions used in Ref. \cite{Dueck:2013gca}, we define the following parameters
\begin{equation}
H\equiv\frac{v^d_{10}}{v_\mathrm{SM}} Y_{10},\quad F\equiv\frac{v^d_{126}}{v_\mathrm{SM}}Y_{126},\quad r\equiv \frac{v^u_{10}}{v^d_{10}}, \quad s\equiv \frac{1}{r}\frac{v^u_{126}}{v^d_{126}} = \frac{v^d_{10}}{v^u_{10}}\frac{v^u_{126}}{v^d_{126}},\quad r_R\equiv v^R_{126}\frac{v_\mathrm{SM}}{v^d_{126}},\quad r_L \equiv  \frac{v_\mathrm{SM}}{v^d_{126}}.
\end{equation}
Using these parameters, we can rewrite Eqs.~\eqref{eq:params}--\eqref{eq:YDelta} as
\begin{equation}\label{eq:matching}
\begin{gathered}
 Y_u = r(H+sF),\quad
Y_d = H+F,\quad
Y_\nu =r(H-3sF),\quad
Y_\ell = H-3F,\\
M_R = r_RF,\quad
Y_\Delta = r_L F.
\end{gathered}
\end{equation}

The $\SO10$ symmetry implies that both $Y_{10}$ and $Y_{126}$ (and hence $H$ and $F$) are complex symmetric matrices. One can choose to work in a basis in which $H$ is diagonal and real. In this basis, $F$ will in general be any complex symmetric matrix. Since $r$, $r_L$, and $r_R$ are just multiplicative factors, their complex phases will have no relevance and they can be chosen to be real. Finally, $s$ remains a complex parameter. 

For type-II seesaw, this parametrization suggests that it is enough to sample $r_L$, or equivalently $v^d_{126}$. However, to perform the matching when integrating out the scalar triplet according to Eq.~\eqref{eq:kappaII}, we also need the value of $v^L_{126}$. Therefore, we sample both $v^L_{126}$ and $v^d_{126}$. Furthermore, we need to sample the mass $M_\Delta$ of the triplet in order to determine at what mass scale the triplet should be integrated out. Note that, although $v^L_{126}$ and $M_\Delta$ are related, there is another parameter in this relation which does not enter elsewhere in our situation. Therefore, we sample $v^L_{126}$ and $M_\Delta$ separately and keep their relationship in mind when deciding the bounds to sample the parameters within.

The total number of parameters for both type-I and type-II seesaw is thus $3(H) + 12(F) + 1(r) + 2(s) + 1(r_R) + 1(v^L_{126}) + 1(v^d_{126}) + 1(M_\Delta) = 22$. If we have type-I seesaw only, we have three parameters less ($v^L_{126}$, $v^d_{126}$, and $M_\Delta$), resulting in 19 parameters, whereas if we have type-II seesaw only, we have one parameter less ($r_R$), resulting in 21 parameters. In principle, the Higgs quartic coupling $\lambda$ at $\MGUT$ should also be included as a parameter in the fit. However, it was observed to consistently be very close to zero and we therefore set $\lambda(\MGUT)=0$ throughout the fits.

There is a constraint on the vevs from the mass of the W boson, which must be fulfilled. That is, they must add in quadrature to the SM Higgs vev. Without type-II seesaw, we have the constraint
\begin{equation}
(v_{10}^u)^2 + (v_{10}^d)^2 + |v_{126}^u|^2 + |v_{126}^d|^2 = v_\mathrm{SM}^2.
\label{eq:normVEV}
\end{equation}
Since the vevs are considered constant in energy, this relation applies at all scales, but is used in our procedure during the sampling at $\MGUT$. Using the definitions of $r$ and $s$, we can rewrite Eq.~\eqref{eq:normVEV} as
\begin{equation}
\left(\frac{v_{10}^d}{v_\mathrm{SM}}\right)^2 \left(1+r^2\right) + \left(\frac{v_{126}^d}{v_\mathrm{SM}}\right)^2 \left(1+r^2s^2\right) = 1.
\end{equation}
Given any parameter values for $r$ and $s$, one can choose $v_{10}^d$ and $v_{126}^d$ such that Eq.~\eqref{eq:normVEV} is satisfied. The only lower bound on the vevs is from perturbativity of the Yukawa couplings. That is, $Y_{10} = v_\mathrm{SM} H / v_{10}^d < \mathcal{O}(1)$ and $Y_{126} = v_\mathrm{SM} F / v_{126}^d < \mathcal{O}(1)$. Although this is usually satisfied in the fits, it should be checked after the parameter values have been obtained.

Including the vev of the scalar triplet, the constraint becomes
\begin{equation}
(v^u_{10})^2 + (v^d_{10})^2 + |v^u_{126}|^2 + |v^d_{126}|^2 + 2(v^L_{126})^2= v_\mathrm{SM}^2,
\end{equation}
where the factor of $2$ comes from the fact that $v^L_{126}$ stems from an $\mathrm{SU}(2)_L$ triplet. Note that in this case, we also need to sample $v_{126}^d$, and thus, we need to make sure that this satisfies any constraints. Since $v_{10}^d$ is still a free parameter and $v^L_{126}$ will be small, we have an absolute limit $\left(v_{126}^d / v_\mathrm{SM}\right)^2 \left(1+r^2s^2\right) < 1$. For the parameter values found, $r = \mathcal{O}(100)$ and $s = \mathcal{O}(0.1)$, this limit implies $|v_{126}^d| \lesssim v_\mathrm{SM}/10$. As mentioned above, these constraints should be checked once the parameter values have been found from the fits. For the parameters related to type-II seesaw, we have experimental bounds $v^L_{126} \lesssim 1\,\mathrm{GeV}$ and $M_\Delta\gtrsim 1\,\mathrm{TeV}$ \cite{Perez:2008ha,Ferreira:2019qpf}.

\subsection{Input Data}\label{sec:ID}

In Tab.~\ref{tab:data}, the 19 input data for the SM observables used in the fits are listed. The masses of the quarks and charged leptons are taken from Ref.~\cite{Deppisch:2018flu}, and the Higgs quartic coupling $\lambda$ is calculated from parameters therein. The neutrino mass-squared differences and the leptonic mixing angles are taken from the global fits presented in Ref.~\cite{deSalas:2017kay}. The CKM parameters have been computed from those listed in the ICHEP 2016 update by the CKMFitter Group~\cite{Charles:2004jd}. For the observable that have a higher precision than $5~\%$, we have chosen to set them to $5~\%$ in order to aid the numerical fitting procedure, as done for example in Ref.~\cite{Dueck:2013gca}. This approach of choice for the errors has the undesired effect of exaggerating the errors on the charged-lepton masses, which are extremely well-known compared to the other observables. However, a very small error on a given observable would mean that any deviation from its central value would cause a large effect on the fit and thereby render the fit almost impossible. The ideal treatment of this issue would be to use the exact values of the charged-lepton masses, as done in some previous fits, see for example Refs.~\cite{Joshipura:2011nn,Altarelli:2013aqa}. However, it is not possible to use such a treatment while solving the RGEs from $M_{\rm GUT}$ to $M_{\rm Z}$ and taking into account the three mass thresholds.

\begin{table}[!ht]
\begin{center}
\begin{tabular}{l l l l}
\hline 
\hline

Observable													& Value & Error 	& Reference	\tabularnewline
\hline
$m_u\,\mathrm{(MeV)}$								&  $1.36$	&  $0.15$ 	&\cite{Deppisch:2018flu}\tabularnewline
$m_c\,\mathrm{(MeV)}$								&  $635$		& $32$ 		&\cite{Deppisch:2018flu}\tabularnewline
$m_t\,\mathrm{(GeV)}$									&  $172$		& $8.7$ 		&\cite{Deppisch:2018flu}\tabularnewline
$m_d\,\mathrm{(MeV)}$								&  $2.90$	& $0.15$		&\cite{Deppisch:2018flu}\tabularnewline
$m_s\,\mathrm{(MeV)}$								&  $54.1$	& $2.8$ 		&\cite{Deppisch:2018flu}\tabularnewline
$m_b\,\mathrm{(GeV)}$								&  $2.87$	& $0.15$		&\cite{Deppisch:2018flu}\tabularnewline

$m_e\,\mathrm{(MeV)}$								&  $0.487$	& $0.025$ 	&\cite{Deppisch:2018flu}\tabularnewline
$m_\mu\,\mathrm{(MeV)}$							&  $103$		& $5.2$ 		&\cite{Deppisch:2018flu}\tabularnewline
$m_\tau\,\mathrm{(GeV)}$							&  $1.75$	& $0.088$ 	&\cite{Deppisch:2018flu}\tabularnewline

$\Delta m_{21}^2\,(10^{-5}\mathrm{eV^2})$	&  $7.55$	&  $0.38$		&\cite{deSalas:2017kay}\tabularnewline
$\Delta m_{31}^2\,(10^{-3}\mathrm{eV^2})$ (NO)	&  $2.50$	&  $0.13$ 	&\cite{deSalas:2017kay}\tabularnewline
$\Delta m_{32}^2\,(10^{-3}\mathrm{eV^2})$ (IO)	&  $-2.42$	&  $0.13$ 	&\cite{deSalas:2017kay}\tabularnewline

$\sin \theta^q_{12}$										&  $0.225$& $0.012$ 	  &\cite{Charles:2004jd}\tabularnewline
$\sin \theta^q_{13}$										&  $0.00372$&$0.00019$  &\cite{Charles:2004jd}\tabularnewline
$\sin \theta^q_{23}$										&  $0.0418$&$0.0021$  &\cite{Charles:2004jd}\tabularnewline
$\delta_\mathrm{CKM}$												&  $1.14$&$0.058$  		  &\cite{Charles:2004jd}\tabularnewline

$\sin^2 \theta^\ell_{12}$								&  $0.320$	&  $0.020$	&\cite{deSalas:2017kay}\tabularnewline
$\sin^2 \theta^\ell_{13}$	(NO)						&  $0.0216$&  $0.0011$	&\cite{deSalas:2017kay}\tabularnewline
$\sin^2 \theta^\ell_{13}$	(IO)						&  $0.0222$&  $0.0012$	&\cite{deSalas:2017kay}\tabularnewline
$\sin^2 \theta^\ell_{23}$	(NO)						&  $0.547$	&  $0.030$ 	&\cite{deSalas:2017kay}\tabularnewline
$\sin^2 \theta^\ell_{23}$	(IO)						&  $0.551$	&  $0.030$ 	&\cite{deSalas:2017kay}\tabularnewline

$\lambda$													&  $0.516$	&  $0.026$ 	&\cite{Deppisch:2018flu}\tabularnewline

\hline 
\hline
\end{tabular}
\caption{\label{tab:data} Data for the SM observables at $M_\mathrm{Z}$ and their corresponding errors used in the fits. The abbreviations NO and IO stand for normal neutrino mass ordering and inverted neutrino mass ordering, respectively.}
\end{center}
\end{table}

\subsection{Numerical Procedure}

The numerical procedure to fit the $\SO10$ parameters to the SM observables consists of two components. One component transforms the $\SO10$ parameters to SM parameters and performs the RG evolution down to $M_\mathrm{Z}$. The other component is a numerical optimization algorithm, which iterates this procedure by sampling different sets of the $\SO10$ parameters with the objective of fitting the derived parameter values to the data.

To relate the $\SO10$ parameters to the fermion observables of the SM, we employ the following procedure:
\begin{enumerate}
\item The required GUT scale parameters are randomly sampled given prior distributions as described below. The number of parameters that are sampled depends on the scenario which is investigated (19 for type-I seesaw, 21 for type-II seesaw, or 22 for type-I+II seesaw).
\item These parameter values are transformed into the parameters of the SM via the matching conditions in Eq.~\eqref{eq:matching}.
\item The SM parameters are evolved from $\MGUT$ down to $M_\mathrm{Z}$ using the RGEs and the matching conditions at each mass threshold. To do this, the following steps are iterated until the parameters have been evolved down all the way to $M_\mathrm{Z}$:
\begin{enumerate}
\item The RGEs for the scenario of interest are used to evolve the parameters down to the first mass threshold, which is either the heaviest right-handed neutrino mass or the scalar triplet mass.
\item At this threshold, the corresponding particle is integrated out following the procedure outlined in Secs.~\ref{sec:typeI} and~\ref{sec:typeII}. This changes the number of parameters, for example by removing some of the parameters associated with the right-handed neutrinos and adding the effective neutrino mass matrix.
\item The RGEs of the new set of parameters are solved to the next threshold.
\end{enumerate}
\item The fermion masses and mixing parameters are calculated from the SM observables at $M_\mathrm{Z}$.
\item These parameter values are compared to the 19 data listed in Tab.~\ref{tab:data} to compute the $\chi^2$ goodness of fit function given by
\begin{equation}\label{eq:chi2}
\chi^2 = \sum_{i=1}^{19} \left(\frac{X_i - \overline{x_i}}{\sigma_i}\right)^2
\end{equation}
in which the current prediction $X_i$ for the $i$th observable is compared to the actual value $\overline{x_i}$ with error $\sigma_i$. Although the $\chi^2$ function usually carries a statistical interpretation, this is non-trivial in problems such as this in which the model is highly non-linear~\cite{Bjorkeroth:2017ybg,Deppisch:2018flu} and may not always be possible.
\end{enumerate}

The above steps are repeated until we converge to a set of parameter values that minimize the $\chi^2$ function. In order to perform the numerical minimization, we link the procedure to the differential evolution algorithm \texttt{Diver} from the \texttt{ScannerBit} package~\cite{Workgroup:2017htr}. We run this parallelized software package on a computing cluster. During this procedure, the parameters are sampled from predetermined distributions with ranges determined from parameter bounds and results of preliminary investigations. For the matrix elements of $H$ and $F$ as well as the vev ratio $r_R$, the vev $v^L_{126}$, and the scalar triplet mass $M_\Delta$, we sample from logarithmic distributions. The parameters $r$, $s$, and $v^d_{126}$ have better known orders of magnitude and are therefore sampled from uniform priors. The parameter ranges are given by
\begin{equation}
\begin{gathered}
H_{11}\in [10^{-7},10^{-5}]_\pm,\quad H_{22}\in [10^{-5},10^{-3}]_\pm,\quad H_{33}\in [10^{-3},10^{-1}]_\pm, \\
|F_{11}| \in [10^{-7},10^{-5}],\quad |F_{12}| \in [10^{-6},10^{-4}], \quad |F_{13}| \in [10^{-5},10^{-3}], \\
|F_{22}| \in [10^{-5},10^{-3}],\quad |F_{23}| \in [10^{-4},10^{-2}], \quad |F_{33}| \in [10^{-4},10^{-2}], \\
r \in [-100,100], \quad |s| \in [0.0,0.5], \quad r_R\in [10^{14},10^{17}], \\
M_\Delta \in [10^6\,\mathrm{GeV},\MGUT], \quad v^L_{126}\in [10^{-9},10^{-1}]\,\mathrm{GeV}, \quad v^d_{126} \in [0.0,10]\,\mathrm{GeV},
\end{gathered}
\end{equation}
where the subscript ``$\pm$" signifies that the parameter is allowed to be positive or negative. The parameters for which the bounds are given as absolute values are complex and their phases are sampled uniformly over $[0,2\pi)$.

After this algorithm has converged, we link the procedure to the basin-hopping algorithm~\cite{bh1997} in the \texttt{SciPy} library~\cite{scipy2001} to further improve the fit, starting from the previously found parameter values. The reason that the algorithms are run in this order is that \texttt{Diver} is more efficient in exploring a large and high-dimensional parameter space, whereas the basin-hopping algorithm improves the fit by perturbing the point in parameter space around the starting point. Finally, we also use a Nelder--Mead simplex algorithm~\cite{Press:1992zz} to further minimize the $\chi^2$ function. 

Using these three algorithms provides some confidence that a reasonable minimum has been found. Note, however, that it is impossible to guarantee that a global minimum has been obtained. To increase our confidence in the minimum found, we run the optimization several times to verify that our set of parameters provides the best fit.

\section{Results and Discussion}\label{sec:results}

The results of the fits show that the known observables of the SM with normal neutrino mass ordering are well accommodated by the model with the type-I seesaw mechanism ($\chi^2 \simeq 14.8$), and that including the type-II seesaw mechanism improves the fit by a small amount to $\chi^2 \simeq 14.7$. The fact that the combination of the two mechanisms provides a better fit than pure type-I seesaw is expected, since it introduces more freedom in the fit. Pure type-II seesaw does not provide as good a fit. With inverted neutrino mass ordering, all fits are much worse than the corresponding ones with normal ordering, in agreement with results of previous fits~\cite{Joshipura:2011nn,Dueck:2013gca,Boucenna:2018wjc} and global fits of neutrino parameters~\cite{deSalas:2017kay}. In Tab.~\ref{tab:chi2}, the resulting values of the $\chi^2$ function in Eq.~\eqref{eq:chi2} are displayed for the six different fits that have been performed. 

\begin{table}[!ht]
\begin{center}
\begin{tabular}{ccccccc}
\hline 
\hline
\multicolumn{3}{c}{Normal ordering} & & \multicolumn{3}{c}{Inverted ordering}\tabularnewline

Type-I+II & Type-I & Type-II & \hspace{0.5cm} & Type-I+II & Type-I & Type-II\tabularnewline
\hline 
14.7 & 14.8 & 119 & & \multicolumn{3}{c}{$\mathcal{O}(1000)$}\tabularnewline
\hline 
\hline
\end{tabular}
\caption{\label{tab:chi2} Values of the $\chi^2$ function for the six different cases considered.}
\end{center}
\end{table}

We find that the best fit with the combination of the type-I and type-II seesaw mechanisms is given by a situation in which the dominant contribution to neutrino mass is given by the type-I seesaw mechanism. This is achieved by having the scalar triplet mass $M_\Delta$ close to $\MGUT$ and a very small value of $v^L_{126}$. The parameter values for the two cases of type-I+II seesaw and pure type-I seesaw, respectively, are the following
\begin{equation}\label{eq:tI+II}
\begin{gathered}
H = 
\begin{pmatrix}
1.00002\times 10^{-7} && 0 && 0 \\
0 && 5.56015\times10^{-5} && 0 \\
0 && 0 && 6.51100\times 10^{-3}
\end{pmatrix}\,,\\
F =
\begin{psmallmatrix}
5.55836\times10^{-6}-3.17854\times 10^{-6}\mathrm{i} && -1.13049\times10^{-5}-1.20803\times 10^{-5}\mathrm{i} && -3.54614\times10^{-5}-1.45941\times 10^{-4}\mathrm{i} \\
-1.13049\times10^{-5}-1.20803\times 10^{-5}\mathrm{i} && -1.63916\times10^{-4}+3.47085\times10^{-5}\mathrm{i} && -2.56495\times10^{-4}+2.55822\times10^{-4}\mathrm{i} \\
-3.54614\times10^{-5}-1.45941\times 10^{-4}\mathrm{i} && -2.56495\times10^{-4}+2.55822\times10^{-4}\mathrm{i} && -9.19624\times10^{-4}-5.02769\times10^{-4}\mathrm{i} \\
\end{psmallmatrix}\,,\\
r = -65.9350, \quad s = 0.391447+0\mathrm{i}, \quad r_R = 2.04454\times10^{15}\,\mathrm{GeV}, \\
M_\Delta = 1.99986\times10^{16}\,\mathrm{GeV}, \quad v^L_{126} = 1.01639\times10^{-6}\,\mathrm{GeV}, \quad v^d_{126} = 4.10530\,\mathrm{GeV}
\end{gathered}
\end{equation}
for ``Type-I+II'' and
\begin{equation}\label{eq:tI}
\begin{gathered}
H = 
\begin{pmatrix}
1.00000\times 10^{-7} && 0 && 0 \\
0 && 5.55975\times10^{-5} && 0 \\
0 && 0 && 6.51243\times 10^{-3}
\end{pmatrix}\,,\\
F =
\begin{psmallmatrix}
5.55892\times10^{-6}-3.17760\times 10^{-6}\mathrm{i} && -1.13026\times10^{-5}-1.20776\times 10^{-5}\mathrm{i} && -3.54522\times10^{-5}-1.45970\times 10^{-4}\mathrm{i} \\
-1.13026\times10^{-5}-1.20776\times 10^{-5}\mathrm{i} && -1.63938\times10^{-4}+3.46916\times10^{-5}\mathrm{i} && -2.56583\times10^{-4}+2.55861\times10^{-4}\mathrm{i} \\
-3.54522\times10^{-5}-1.45970\times 10^{-4}\mathrm{i} && -2.56583\times10^{-4}+2.55861\times10^{-4}\mathrm{i} && -9.19847\times10^{-4}-5.02620\times10^{-4}\mathrm{i} \\
\end{psmallmatrix}\,,\\
r = -66.0472, \quad s = 0.391339+0\mathrm{i}, \quad r_R = 2.05408\times10^{15}\,\mathrm{GeV}
\end{gathered}
\end{equation}
for ``Type-I''. Note that although $s$ is a complex parameter, the fit favored a value with a negligibly small imaginary part. As can be observed, the parameter values are very close to each other in the two cases. In fact, the parameter values for type-I+II seesaw were found by starting from the parameter values for pure type-I seesaw and adding a very small contribution from the scalar triplet $\Delta$, i.e.~a large $M_\Delta$ and a small $v^L_{126}$, and using the basin-hopping and simplex algorithms to improve the parameter values. This biased procedure was found to produce a better fit than the one which started from unknown values of the parameters. Hence, the results of the two fits suggest that the best fit is provided by a set of parameter values for which the type-I seesaw mechanism is the dominant contribution to neutrino mass, but the type-II seesaw mechanism has a small contribution.

\begin{figure}[t]
\includegraphics[clip,width=0.8\textwidth]{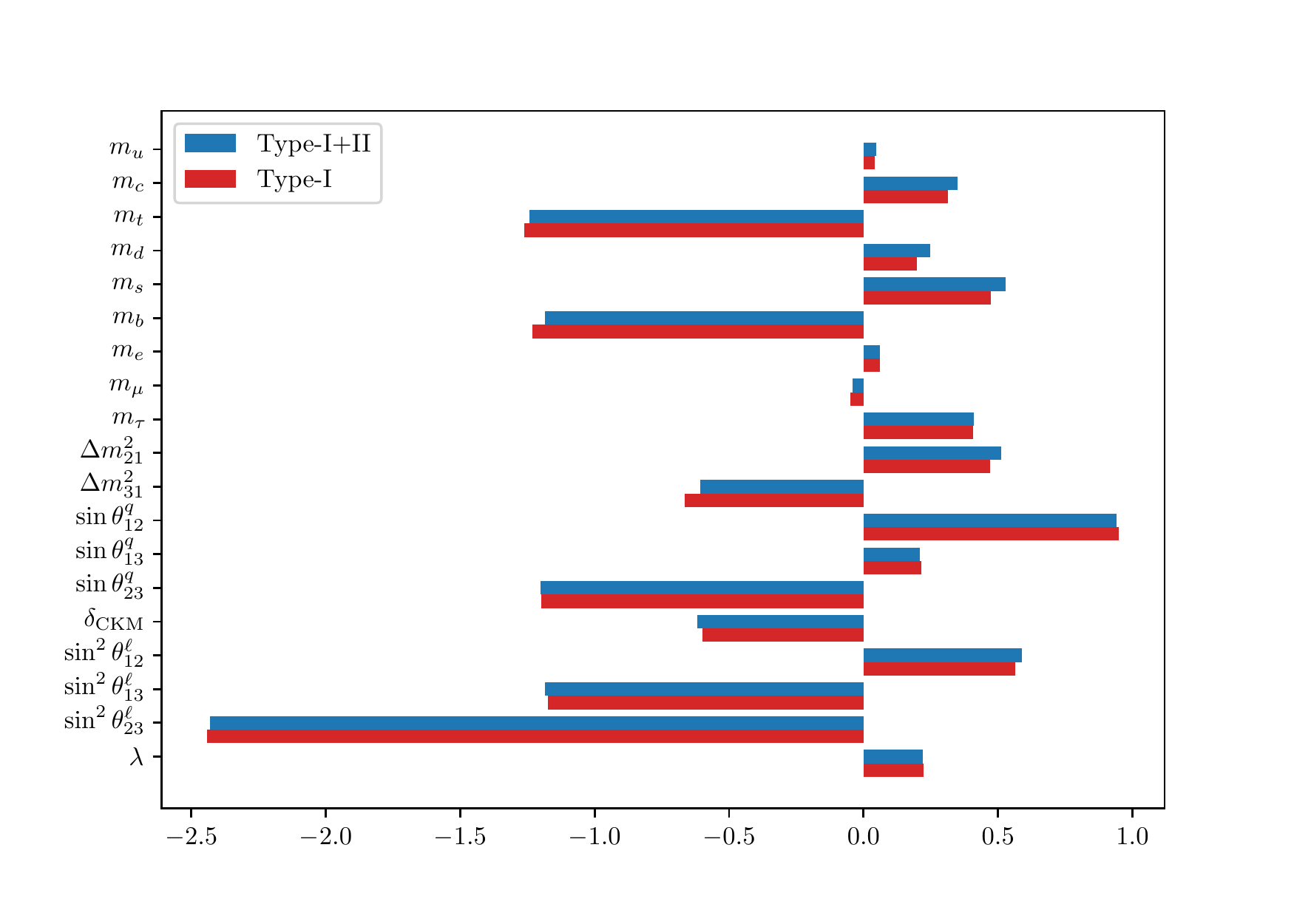}
\caption{\label{fig:pulls} Pulls corresponding to the SM observables for the two cases of type-I+II seesaw and pure type-I seesaw, respectively, and normal neutrino mass ordering. The sum of the pulls squared gives the $\chi^2$ value.}
\end{figure}

In Tab.~\ref{tab:predictions}, the resulting values of the observables for the ``Type-I+II'' parameter values in Eq.~\eqref{eq:tI+II} and the ``Type-I'' parameter values in Eq.~\eqref{eq:tI} are shown. The corresponding pulls are also shown, which are the quantities whose squares are summed to give the value of the $\chi^2$ function in Eq.~\eqref{eq:chi2}. In Fig.~\ref{fig:pulls}, these pulls are further displayed for ease of comparison. Since the parameter values and $\chi^2$ values are close in the two cases, their pulls are also similar. The largest contribution to the $\chi^2$ values comes from $\sin^2\theta^\ell_{23}$, for which a lower value than the measured one is predicted. In fact, the fit results suggest a value of $\theta^\ell_{23}$ in the lower octant, whereas the data favors a value in the higher octant~\cite{deSalas:2017kay}. This tension in the values of $\sin^2\theta^\ell_{23}$ has been observed in similar fits to ours~\cite{Dueck:2013gca,Boucenna:2018wjc}. If future neutrino experiments shift the value of $\theta^\ell_{23}$ to be in the lower octant, the goodness of the fits presented in this work would be greatly improved. In fact, the octant of $\theta^\ell_{23}$ the lower octant is allowed at $1\sigma$ in one of the global fits to neutrino data~\cite{Capozzi:2018ubv} and consistently allowed at $3\sigma$~\cite{deSalas:2017kay,Capozzi:2018ubv,Esteban:2018azc,nu-fit18}.

\begin{table}[!htbp]
\begin{center}
\begin{tabular}{l  l l  l l}
\hline 
\hline
Observable/ & \multicolumn{2}{c}{Type-I+II} & \multicolumn{2}{c}{Type-I}  \tabularnewline
Parameter\\
	& Value & Pull & Value & Pull \tabularnewline

\hline 

$m_u$				& $1.37\,\mathrm{MeV}$ 	& $0.0473$	& 	$1.37\,\mathrm{MeV}$&$0.0422$ \tabularnewline
$m_c$				& $646\,\mathrm{MeV}$	& $0.351$		& 	$645\,\mathrm{MeV}$&$0.314$ \tabularnewline
$m_t$				& $161\,\mathrm{GeV}$	& $-1.24$		&$161\,\mathrm{GeV}$	& $-1.26$	  \tabularnewline
$m_d$				& $2.94\,\mathrm{MeV}$	&$0.248$		& $2.93\,\mathrm{MeV}$& $0.199$\tabularnewline
$m_s$				& $55.6\,\mathrm{MeV}$	&$0.528$		& $55.4\,\mathrm{MeV}$&$0.473$ \tabularnewline
$m_b$				& $2.69\,\mathrm{GeV}$	& $-1.18$		& $2.69\,\mathrm{GeV}$&$-1.23$ \tabularnewline
$m_e$				& $0.489\,\mathrm{MeV}$& $0.0613$	& $0.489\,\mathrm{MeV}$&$0.0613$ \tabularnewline
$m_\mu$  			& $103\,\mathrm{MeV}$	& $-0.0421$	&	$103\,\mathrm{MeV}$&$-0.0482$ \tabularnewline
$m_\tau$  		& $1.79\,\mathrm{GeV}$	& $0.411$		&	$1.79\,\mathrm{GeV}$&$0.408$ \tabularnewline

$\Delta m_{21}^2$	& $7.74\times10^{-5}\,\mathrm{eV}^2$&$0.511$& $7.73\times10^{-5}\,\mathrm{eV}^2$ &$0.470$ \tabularnewline
$\Delta m_{31}^2$	& $2.42\times10^{-3}\,\mathrm{eV}^2$&$-0.608$& $2.41\times10^{-3}\,\mathrm{eV}^2$&$-0.665$ \tabularnewline

$\sin \theta^q_{12}$			& $0.236$	&$0.942$		& 	$0.236$&$0.950$ \tabularnewline
$\sin \theta^q_{13}$			& $0.00376$&  $0.210$	& $0.00376$&$0.215$ \tabularnewline
$\sin \theta^q_{23}$			& $0.0393$	&$-1.20$		&$0.0393$ & $-1.20$ \tabularnewline
$\delta_\mathrm{CKM}$	& $1.10$		& $-0.618$	& $1.11$ &$-0.598$ \tabularnewline

$\sin^2 \theta^\ell_{12}$	& $0.332$	& $0.589$	& $0.331$ & $0.565$ \tabularnewline
$\sin^2 \theta^\ell_{13}$	& $0.0203$	& $-1.18$		&$0.0203$ & $-1.17$\tabularnewline
$\sin^2 \theta^\ell_{23}$	& $0.474$	& $-2.43$		& $0.474$& $-2.44$ \tabularnewline

$\lambda$						& $0.522$	& $0.221$		& $0.522$ & $0.222$ \tabularnewline
\hline
$\chi^2$							&  	& $14.7$   	& 	& $14.8$ \tabularnewline
\hline
$m_1$ 								&  $3.70\times10^{-3}\,\mathrm{eV}$&   --& 	$3.70\times10^{-3}\,\mathrm{eV}$& --\tabularnewline
$m_2$ 								& $9.55\times10^{-3}\,\mathrm{eV}$&   --& 	$9.54\times10^{-3}\,\mathrm{eV}$& --\tabularnewline
$m_3$					 			&  $4.93\times10^{-2}\,\mathrm{eV}$&   --& 	$4.93\times10^{-2}\,\mathrm{eV}$& --\tabularnewline
$M_1$ 								&$1.87\times10^{10}\,\mathrm{GeV}$&   --& 	$1.88\times10^{10}\,\mathrm{GeV}$& --\tabularnewline
$M_2$						 		&$4.46\times10^{11}\,\mathrm{GeV}$&   --& $4.48\times10^{11}\,\mathrm{GeV}$& --\tabularnewline
$M_3$						 		&$2.34\times10^{12}\,\mathrm{GeV}$&   --& $2.36\times10^{12}\,\mathrm{GeV}$& --\tabularnewline
$m_{ee}$						 	&$1.56\times10^{-3}\,\mathrm{eV}$	&   --& $1.56\times10^{-3}\,\mathrm{eV}$& --\tabularnewline
$\delta_\mathrm{CP}$ 		&$0.441$											&   --& $0.447$ & --\tabularnewline
\hline 
\hline
\end{tabular}
\caption{\label{tab:predictions} Predicted values of the observables that are included in the fits together with the associated pulls for the two cases of type-I+II seesaw and pure type-I seesaw, respectively, and normal neutrino mass ordering. Shown are also predicted values for some of the unknown parameters of the neutrino sector, which are not included in the fits, namely the small neutrino masses $m_1$, $m_2$, and $m_3$, the large right-handed neutrino masses $M_1$, $M_2$, and $M_3$, the effective neutrinoless double beta decay mass $m_{ee}$, and the leptonic CP-violating phase $\delta_\mathrm{CP}$.}
\end{center}
\end{table}

Also shown in Tab.~\ref{tab:predictions} are predictions for some parameters not included in the fits, namely the small neutrino masses $m_1$, $m_2$, and $m_3$, the large right-handed neutrino masses $M_1$, $M_2$, and $M_3$, the effective neutrinoless double beta decay mass $m_{ee}$, and the leptonic CP-violating phase $\delta_\mathrm{CP}$. From the small neutrino masses, we note that their sum is below the cosmological upper limit~\cite{Tanabashi:2018oca}, and further that $m_{ee}$ is within the allowed region~\cite{Deppisch:2014zta,Pas:2015eia}. As for $\delta_\mathrm{CP}$, the predicted value is far from the value favored by global fits~\cite{deSalas:2017kay}. Although this seems like a failure of the fits, one should keep in mind that it is possible that including it as an observable in the fits may still yield an acceptable value and further that its exact value has not been directly measured.

\begin{figure*}[t]
\begin{tabular}{c c}
\includegraphics[width=0.495\textwidth]{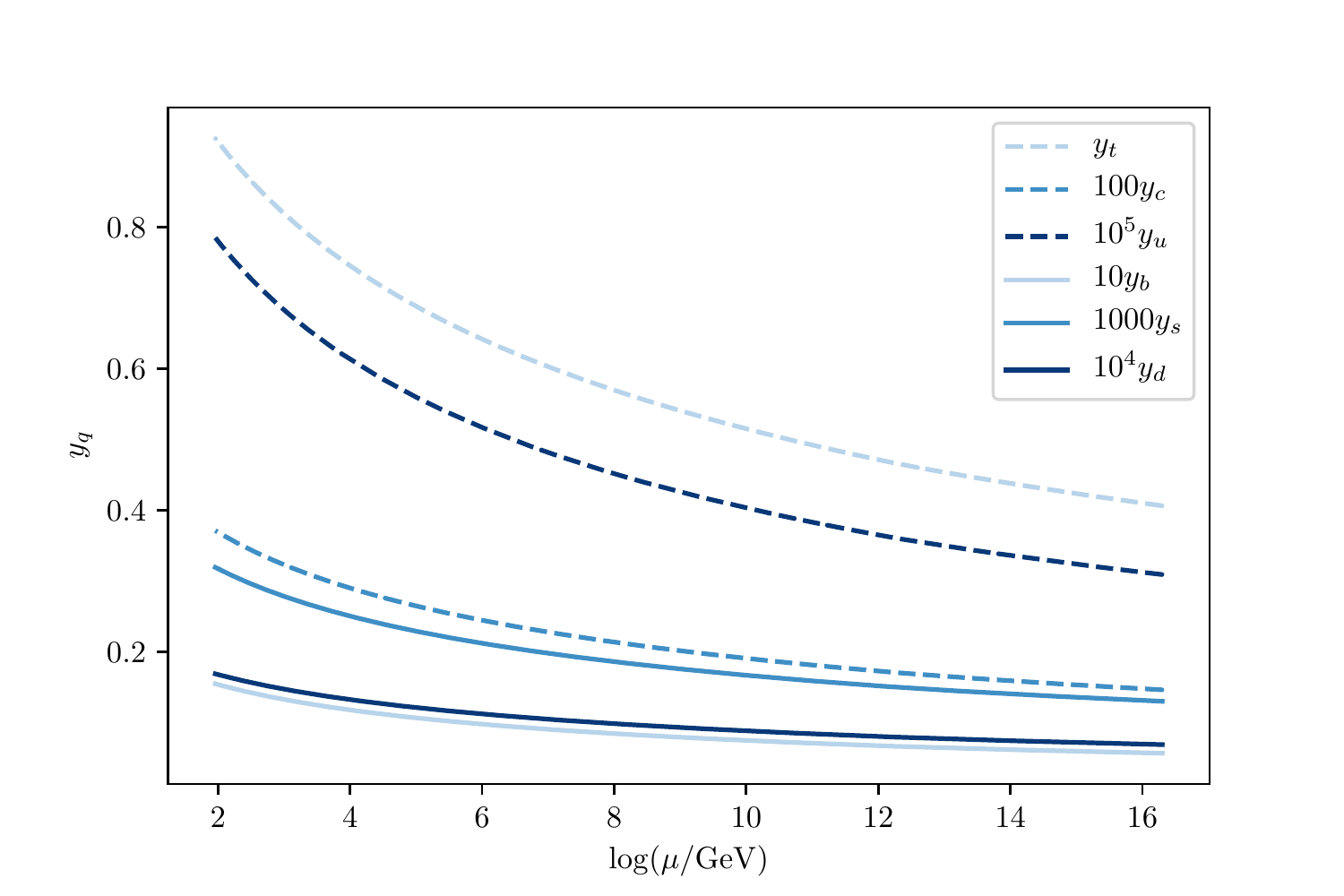} &

\includegraphics[width=0.495\textwidth]{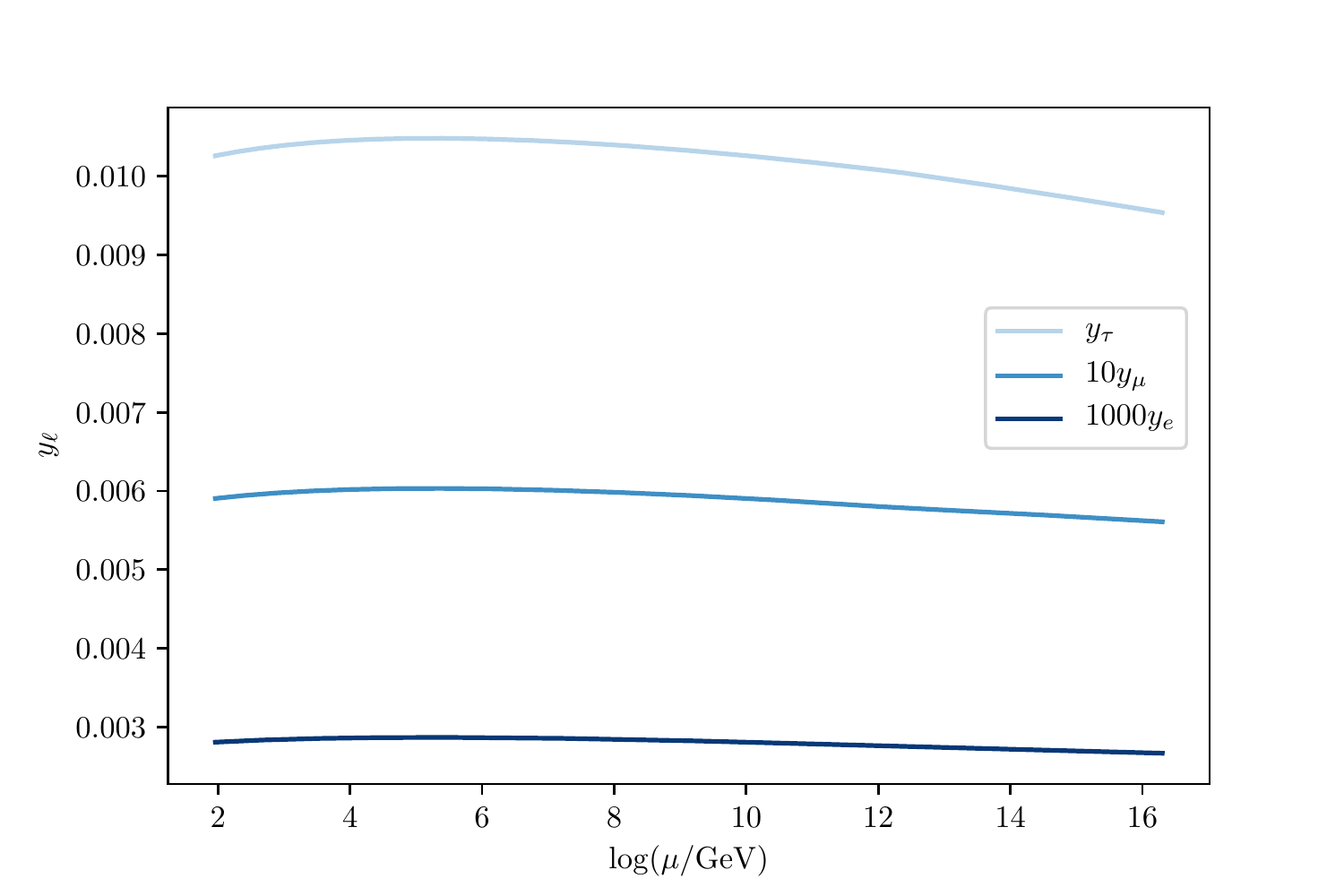}\\

\includegraphics[width=0.495\textwidth]{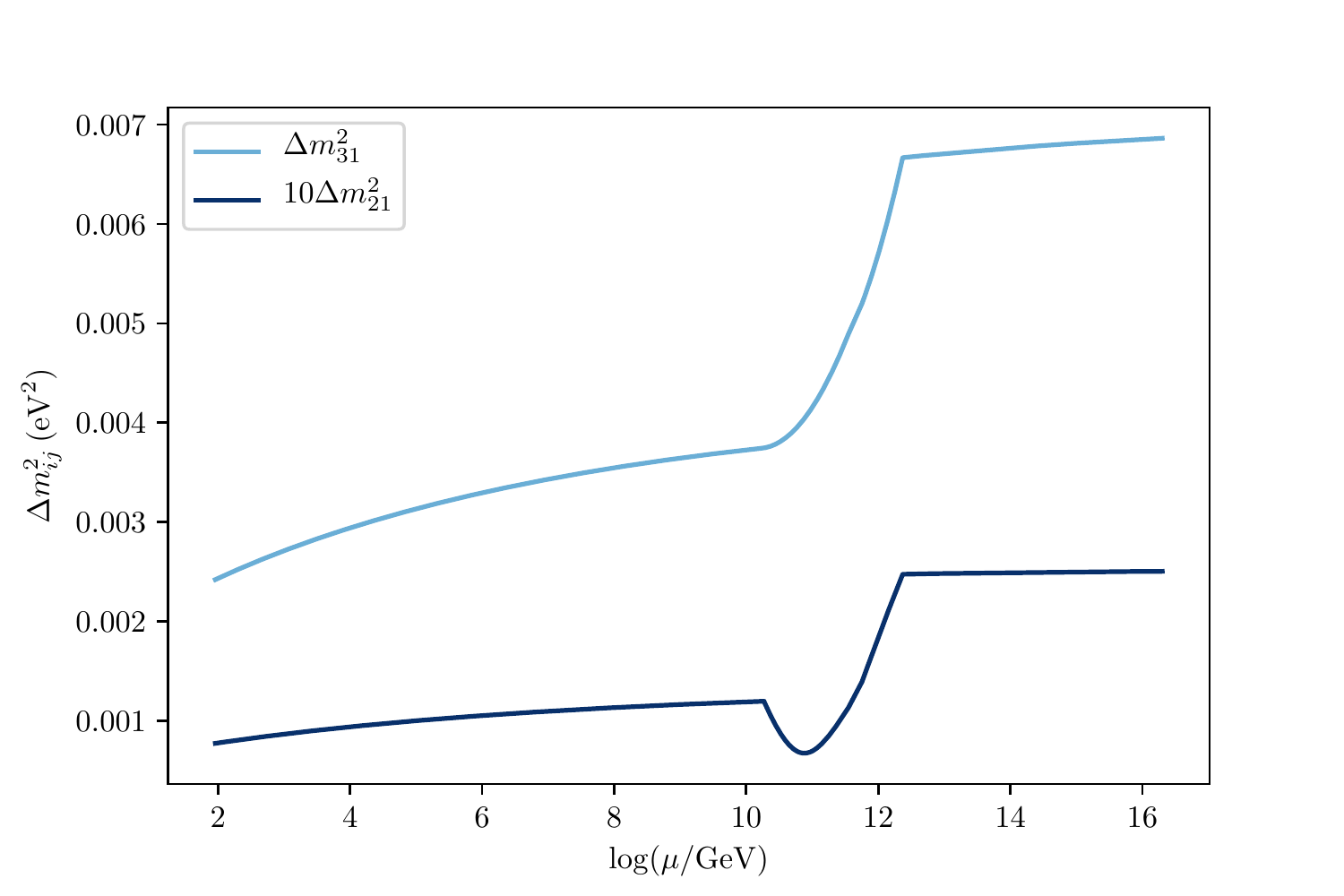}&

\includegraphics[width=0.495\textwidth]{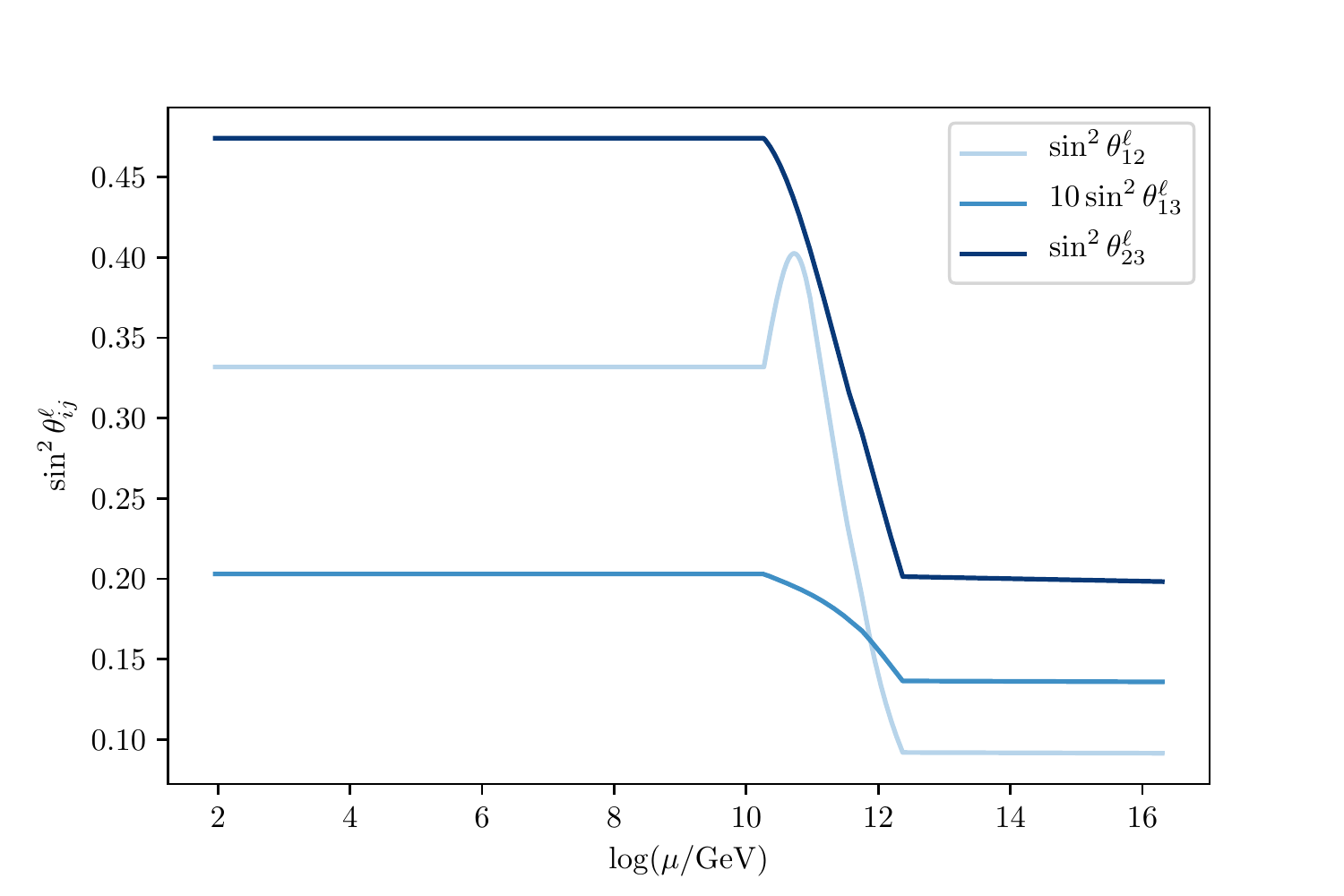}
\end{tabular}

\caption{\label{fig:running}Renormalization group evolution of some quantities for the fit with type-I+II seesaw and normal neutrino mass ordering as functions of the energy scale $\mu$ between $M_\mathrm{Z}$ and $\MGUT$. \emph{Upper-left panel:} Quark Yukawa couplings (singular values of $Y_u$ and $Y_d$). \emph{Upper-right panel:} Charged-lepton Yukawa couplings (singular values of $Y_\ell$). \emph{Lower-left panel:} Neutrino mass-squared differences (Differences of squared singular values of the neutrino mass matrix, multiplied by $v_\mathrm{SM}^2/4$). \emph{Lower-right panel:} Sine squareds of the leptonic mixing angles calculated from the mixing between the charged-lepton Yukawa matrix and the neutrino mass matrix.}
\end{figure*}

In Fig.~\ref{fig:running}, the RG evolution for some of the quantities are displayed using the parameter values in Eq.~\eqref{eq:tI+II}. First, the upper-left panel shows the RG evolution of the quark Yukawa couplings. These are the singular values of the Yukawa matrices $Y_u$ and $Y_d$. Second, the upper-right panel shows the charged-lepton Yukawa couplings, calculated in the same way from the Yukawa matrix $Y_\ell$. Third, the lower-left panel shows the RG evolution of the neutrino mass-squared differences, calculated from the differences of the squared singular values of the neutrino mass matrix. This is taken as $\kappa$ plus the type-I seesaw contributions from the neutrino Yukawa couplings that have not yet been integrated out. Finally, the lower-right panel shows the RG evolution of the leptonic mixing angles, which are calculated by computing the leptonic mixing matrix as the mixing matrix between the charged-lepton Yukawa matrix and the neutrino mass matrix. 

As can be seen in the two lower panels, the right-handed neutrino mass thresholds considerably affect the RG evolution of the parameters related to neutrino masses. This enhanced RG evolution between $M_1$ and $M_3$ can be understood by noting that in order to derive these quantities, one must consider $\kappa_\mathrm{eff}=\kappa+2Y_\nu^TM^{-1}Y_\nu$, since some but not all right-handed neutrinos have been integrated out in this energy region. The RG evolution of the neutrino mass-squared differences are thus due to the RG evolution of both terms in $\kappa_\mathrm{eff}$. This causes an enhancement of the RG evolution inside the energy region compared to outside it, since its RG evolution is given by
\begin{equation}\label{eq:kappa_eff}
\frac{d\kappa_\mathrm{eff}}{d \ln\mu} = \frac{d\kappa}{d \ln\mu} + 2 \frac{dY_\nu^T}{d \ln\mu}M^{-1}Y_\nu + 2 Y_\nu^T M^{-1} \frac{dY_\nu}{d \ln\mu} - 2Y_\nu^T M^{-1} \frac{dM}{d\ln\mu}M^{-1} Y_\nu.
\end{equation}
Between $M_1$ and $M_3$, the second and third terms in the right-hand side of Eq.~\eqref{eq:kappa_eff} are about three to five larger than the first term. Below $M_1$, only the first term contributes. Its contribution is smaller than above $M_1$ due to the absence of the terms with $Y_\nu$ in Eq.~\eqref{eq:kappaRGE}, since $Y_\nu$ is not a parameter of the theory after the right-handed neutrinos have been integrated out. Above $M_3$, only the last three terms in the right-hand side of Eq.~\eqref{eq:kappa_eff} contribute, but they are suppressed by $M_3$, while below $M_3$ they are only suppressed by $M_2$. Furthermore, note that the parameters related to neutrino masses (i.e.~the neutrino mass-squared differences and the leptonic mixing angles) should only be considered effective parameters between $M_1$ and $M_3$, i.e.~before the actual light neutrino mass matrix is formed.

\begin{figure}
\includegraphics[clip,width=0.8\textwidth]{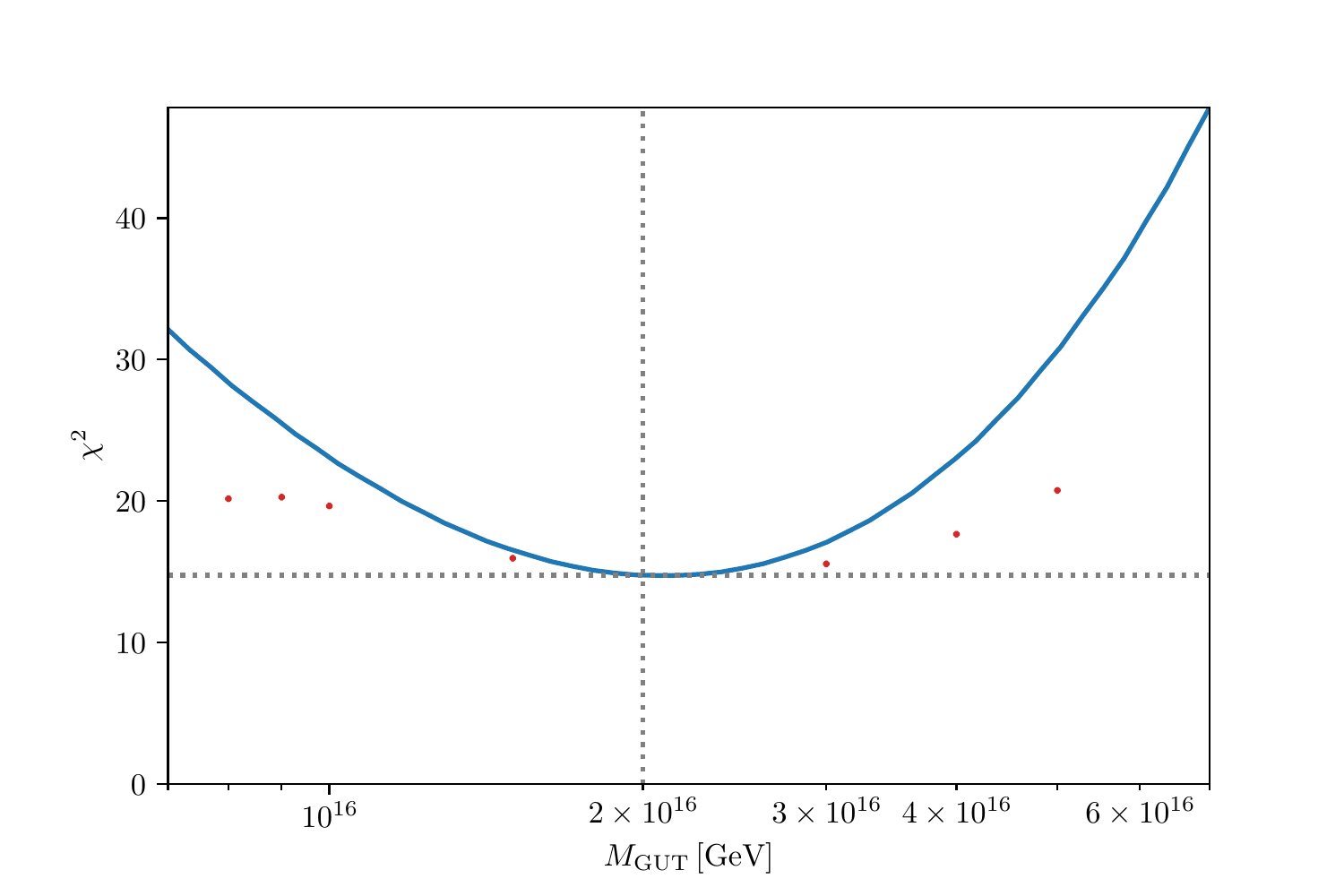}
\caption{\label{fig:MGUT} Variation of the $\chi^2$ function with $\MGUT$. The blue curve shows the result of varying $\MGUT$, while keeping the same best-fit parameter values. The red dots show the result of performing local optimization around the best-fit parameter values for some selected values of $\MGUT$.}
\end{figure}

Since the model presented is not a complete $\SO10$ model in the sense that we have not imposed gauge coupling unification, it is relevant to consider the effect on the $\chi^2$ function by changing the value of $\MGUT$. We assume that gauge coupling unification is taken care of by some new physics between $M_\mathrm{Z}$ and $\MGUT$. Such a model can be found in e.g.~Refs.~\cite{Frigerio:2009wf,Parida:2016hln,Boucenna:2018wjc}. Our aim in this work is to present general results of fits for the seesaw mechanisms of types I and II, for which we need to verify that the results are not sensitive to the exact unification scale. Indeed, we find that the $\chi^2$ value is fairly insensitive to changes in $\MGUT$, as shown in Fig.~\ref{fig:MGUT}. The blue curve shows the value of the $\chi^2$ function for various values of $\MGUT$ in the type-I+II seesaw case with the parameter values of Eq.~\eqref{eq:tI+II}. These $\chi^2$ values will, of course, not be the ones found from fits performed with the given $\MGUT$. After running the basin-hopping algorithm starting from those parameter values, the points shown in red were found. 

Thus, one can conclude that if a fit was performed with a given $\MGUT$, the resulting $\chi^2$ values would be at most the ones given by the red dots. Since these are very close to the $\chi^2$ value corresponding to our choice of $\MGUT$, one can conclude that the ability to find an acceptable fit applies also to realistic models.\footnote{This neglects the fact that changes in $\MGUT$ will in general be accompanied by changes in the RG evolution of the Yukawa couplings due to new physics between $M_\mathrm{Z}$ and $\MGUT$.} The results should not be interpreted to suggest that the value $\MGUT=2\times 10^{16}\,\mathrm{GeV}$ provides the best fit. The reason why this has the lowest $\chi^2$ value is simply that the original fit was performed with that value and the other points were found by perturbing the corresponding parameter values. 

\section{Summary and Conclusions}\label{sec:summary}

The parameters of minimal non-SUSY $\SO10$ models have been fitted to the known observables of the SM with neutrino mass generated via type-I or type-II seesaw, or a combination thereof. We have refrained from specifying a complete model by not imposing gauge coupling unification, which we have assumed can be taken care of by some new physics between $M_\mathrm{Z}$ and $\MGUT$. As opposed to many similar works, we have performed the fits by sampling the parameters of the $\SO10$ model, and evolving them down to $M_\mathrm{Z}$ using the RGEs, where the corresponding parameter values have been compared to the known values of the observables. This method allows for a proper treatment of mass thresholds due to the masses of the right-handed neutrinos. Furthermore, we have used updated data at $M_\mathrm{Z}$.

The results have shown that it is possible to find acceptable fits only in the case of normal neutrino mass ordering, in agreement with hints from global fits to neutrino data. The combination of the seesaw mechanisms of types I and II provides the best fit, which is marginally better than that with the type-I seesaw mechanism only. This, together with the parameter values of the best fit, suggests that the type-I seesaw mechanism is the dominant contributor to neutrino mass. 

From the values of best-fit parameters, it can be observed that the largest contribution to the $\chi^2$ function is coming from $\sin^2\theta^\ell_{23}$, which is favored to be below $0.5$ by our fits, thus predicting the leptonic mixing angle $\theta^\ell_{23}$ in the lower octant in contrast to the latest neutrino data. The best-fit parameters also provide predictions on some of the unknown neutrino parameters, such as the absolute neutrino masses, the effective neutrinoless double beta decay mass, and the leptonic CP-violating phase.

Finally, to verify the robustness of these fits to model-dependent changes of the unification scale, we have investigated the effect of varying $\MGUT$ around our chosen value of $2\times 10^{16}\,\mathrm{GeV}$. After performing local minimization around our best-fit parameter values, we have found that the effect of such variations of $\MGUT$ is small. Therefore, the results presented can be considered to be somewhat general for minimal non-SUSY $\SO10$ models, and thus, we expect the fits of specific models, which impose gauge coupling unification by new low-scale physics, to be similar.

In general, future experiments will continue to improve the precision of the observables of the SM, which will affect future fits of the kind presented in this paper. In particular, the value of $\theta^\ell_{23}$ has been shown to have a large effect on the fits. Determination of the effective neutrinoless double beta decay mass as well as the leptonic CP-violating phase will also provide invaluable information to determine the ability of different models to reproduce the observables of the SM.

\begin{acknowledgments}
The authors wish to thank Sofiane M.~Boucenna for collaboration in early stages of this project. T.O.~acknowledges support by the Swedish Research Council (Vetenskapsrådet) through contract No.~2017-03934 and the KTH Royal Institute of Technology for a sabbatical period at the University of Iceland. M.P.~thanks “Stiftelsen Olle Engkvist Byggmästare” and “Roland Gustafssons Stiftelse för teoretisk fysik” for financial support. Numerical computations were performed on resources provided by the Swedish National Infrastructure for Computing (SNIC) at PDC Center for High Performance Computing (PDC-HPC) at KTH Royal Institute of Technology in Stockholm, Sweden under project numbers PDC-2018-49 and SNIC 2018/3-559.
\end{acknowledgments}

\bibliographystyle{apsrev4-1}
\bibliography{refs.bib}

\begin{thebibliography}{58}%
\makeatletter
\providecommand \@ifxundefined [1]{%
 \@ifx{#1\undefined}
}%
\providecommand \@ifnum [1]{%
 \ifnum #1\expandafter \@firstoftwo
 \else \expandafter \@secondoftwo
 \fi
}%
\providecommand \@ifx [1]{%
 \ifx #1\expandafter \@firstoftwo
 \else \expandafter \@secondoftwo
 \fi
}%
\providecommand \natexlab [1]{#1}%
\providecommand \enquote  [1]{``#1''}%
\providecommand \bibnamefont  [1]{#1}%
\providecommand \bibfnamefont [1]{#1}%
\providecommand \citenamefont [1]{#1}%
\providecommand \href@noop [0]{\@secondoftwo}%
\providecommand \href [0]{\begingroup \@sanitize@url \@href}%
\providecommand \@href[1]{\@@startlink{#1}\@@href}%
\providecommand \@@href[1]{\endgroup#1\@@endlink}%
\providecommand \@sanitize@url [0]{\catcode `\\12\catcode `\$12\catcode
  `\&12\catcode `\#12\catcode `\^12\catcode `\_12\catcode `\%12\relax}%
\providecommand \@@startlink[1]{}%
\providecommand \@@endlink[0]{}%
\providecommand \url  [0]{\begingroup\@sanitize@url \@url }%
\providecommand \@url [1]{\endgroup\@href {#1}{\urlprefix }}%
\providecommand \urlprefix  [0]{URL }%
\providecommand \Eprint [0]{\href }%
\providecommand \doibase [0]{http://dx.doi.org/}%
\providecommand \selectlanguage [0]{\@gobble}%
\providecommand \bibinfo  [0]{\@secondoftwo}%
\providecommand \bibfield  [0]{\@secondoftwo}%
\providecommand \translation [1]{[#1]}%
\providecommand \BibitemOpen [0]{}%
\providecommand \bibitemStop [0]{}%
\providecommand \bibitemNoStop [0]{.\EOS\space}%
\providecommand \EOS [0]{\spacefactor3000\relax}%
\providecommand \BibitemShut  [1]{\csname bibitem#1\endcsname}%
\let\auto@bib@innerbib\@empty
\bibitem [{\citenamefont {Georgi}\ and\ \citenamefont
  {Glashow}(1974)}]{Georgi:1974sy}%
  \BibitemOpen
  \bibfield  {author} {\bibinfo {author} {\bibfnamefont {H.}~\bibnamefont
  {Georgi}}\ and\ \bibinfo {author} {\bibfnamefont {S.~L.}\ \bibnamefont
  {Glashow}},\ }\href {\doibase 10.1103/PhysRevLett.32.438} {\bibfield
  {journal} {\bibinfo  {journal} {Phys. Rev. Lett.}\ }\textbf {\bibinfo
  {volume} {32}},\ \bibinfo {pages} {438} (\bibinfo {year} {1974})}\BibitemShut
  {NoStop}%
\bibitem [{\citenamefont {Fritzsch}\ and\ \citenamefont
  {Minkowski}(1975)}]{Fritzsch:1974nn}%
  \BibitemOpen
  \bibfield  {author} {\bibinfo {author} {\bibfnamefont {H.}~\bibnamefont
  {Fritzsch}}\ and\ \bibinfo {author} {\bibfnamefont {P.}~\bibnamefont
  {Minkowski}},\ }\href {\doibase 10.1016/0003-4916(75)90211-0} {\bibfield
  {journal} {\bibinfo  {journal} {Annals Phys.}\ }\textbf {\bibinfo {volume}
  {93}},\ \bibinfo {pages} {193} (\bibinfo {year} {1975})}\BibitemShut
  {NoStop}%
\bibitem [{\citenamefont {Minkowski}(1977)}]{Minkowski:1977sc}%
  \BibitemOpen
  \bibfield  {author} {\bibinfo {author} {\bibfnamefont {P.}~\bibnamefont
  {Minkowski}},\ }\href {\doibase 10.1016/0370-2693(77)90435-X} {\bibfield
  {journal} {\bibinfo  {journal} {Phys. Lett. B}\ }\textbf {\bibinfo {volume}
  {67}},\ \bibinfo {pages} {421} (\bibinfo {year} {1977})}\BibitemShut
  {NoStop}%
\bibitem [{\citenamefont {Gell-Mann}\ \emph {et~al.}(1979)\citenamefont
  {Gell-Mann}, \citenamefont {Ramond},\ and\ \citenamefont
  {Slansky}}]{GellMann:1980vs}%
  \BibitemOpen
  \bibfield  {author} {\bibinfo {author} {\bibfnamefont {M.}~\bibnamefont
  {Gell-Mann}}, \bibinfo {author} {\bibfnamefont {P.}~\bibnamefont {Ramond}}, \
  and\ \bibinfo {author} {\bibfnamefont {R.}~\bibnamefont {Slansky}},\
  }\bibfield  {booktitle} {\emph {\bibinfo {booktitle} {{Supergravity Workshop
  Stony Brook, New York, September 27-28, 1979}}},\ }\href@noop {} {\bibfield
  {journal} {\bibinfo  {journal} {Conf. Proc.}\ }\textbf {\bibinfo {volume}
  {C790927}},\ \bibinfo {pages} {315} (\bibinfo {year} {1979})},\ \Eprint
  {http://arxiv.org/abs/1306.4669}{arXiv:1306.4669 [hep-th]}\BibitemShut
  {NoStop}%
\bibitem [{\citenamefont {Mohapatra}\ and\ \citenamefont
  {Senjanovi{\'c}}(1980)}]{Mohapatra:1979ia}%
  \BibitemOpen
  \bibfield  {author} {\bibinfo {author} {\bibfnamefont {R.~N.}\ \bibnamefont
  {Mohapatra}}\ and\ \bibinfo {author} {\bibfnamefont {G.}~\bibnamefont
  {Senjanovi{\'c}}},\ }\href {\doibase 10.1103/PhysRevLett.44.912} {\bibfield
  {journal} {\bibinfo  {journal} {Phys. Rev. Lett.}\ }\textbf {\bibinfo
  {volume} {44}},\ \bibinfo {pages} {912} (\bibinfo {year} {1980})}\BibitemShut
  {NoStop}%
\bibitem [{\citenamefont {Yanagida}(1979)}]{Yanagida:1979}%
  \BibitemOpen
  \bibfield  {author} {\bibinfo {author} {\bibfnamefont {T.}~\bibnamefont
  {Yanagida}},\ }\bibfield  {booktitle} {\emph {\bibinfo {booktitle}
  {{Proceedings: Workshop on the Unified Theories and the Baryon Number in the
  Universe: Tsukuba, Japan, February 13-14, 1979}}},\ }\href@noop {} {\bibfield
   {journal} {\bibinfo  {journal} {Conf. Proc.}\ }\textbf {\bibinfo {volume}
  {C7902131}},\ \bibinfo {pages} {95} (\bibinfo {year} {1979})}\BibitemShut
  {NoStop}%
\bibitem [{\citenamefont {Schechter}\ and\ \citenamefont
  {Valle}(1980)}]{Schechter:1980gr}%
  \BibitemOpen
  \bibfield  {author} {\bibinfo {author} {\bibfnamefont {J.}~\bibnamefont
  {Schechter}}\ and\ \bibinfo {author} {\bibfnamefont {J.~W.~F.}\ \bibnamefont
  {Valle}},\ }\href {\doibase 10.1103/PhysRevD.22.2227} {\bibfield  {journal}
  {\bibinfo  {journal} {Phys. Rev. D}\ }\textbf {\bibinfo {volume} {22}},\
  \bibinfo {pages} {2227} (\bibinfo {year} {1980})}\BibitemShut {NoStop}%
\bibitem [{\citenamefont {Magg}\ and\ \citenamefont
  {Wetterich}(1980)}]{Magg:1980ut}%
  \BibitemOpen
  \bibfield  {author} {\bibinfo {author} {\bibfnamefont {M.}~\bibnamefont
  {Magg}}\ and\ \bibinfo {author} {\bibfnamefont {C.}~\bibnamefont
  {Wetterich}},\ }\href {\doibase 10.1016/0370-2693(80)90825-4} {\bibfield
  {journal} {\bibinfo  {journal} {Phys. Lett. B}\ }\textbf {\bibinfo {volume}
  {94}},\ \bibinfo {pages} {61} (\bibinfo {year} {1980})}\BibitemShut {NoStop}%
\bibitem [{\citenamefont {Lazarides}\ \emph {et~al.}(1981)\citenamefont
  {Lazarides}, \citenamefont {Shafi},\ and\ \citenamefont
  {Wetterich}}]{Lazarides:1980nt}%
  \BibitemOpen
  \bibfield  {author} {\bibinfo {author} {\bibfnamefont {G.}~\bibnamefont
  {Lazarides}}, \bibinfo {author} {\bibfnamefont {Q.}~\bibnamefont {Shafi}}, \
  and\ \bibinfo {author} {\bibfnamefont {C.}~\bibnamefont {Wetterich}},\ }\href
  {\doibase 10.1016/0550-3213(81)90354-0} {\bibfield  {journal} {\bibinfo
  {journal} {Nucl. Phys. B}\ }\textbf {\bibinfo {volume} {181}},\ \bibinfo
  {pages} {287} (\bibinfo {year} {1981})}\BibitemShut {NoStop}%
\bibitem [{\citenamefont {Mohapatra}\ and\ \citenamefont
  {Senjanovi{\'c}}(1981)}]{Mohapatra:1980yp}%
  \BibitemOpen
  \bibfield  {author} {\bibinfo {author} {\bibfnamefont {R.~N.}\ \bibnamefont
  {Mohapatra}}\ and\ \bibinfo {author} {\bibfnamefont {G.}~\bibnamefont
  {Senjanovi{\'c}}},\ }\href {\doibase 10.1103/PhysRevD.23.165} {\bibfield
  {journal} {\bibinfo  {journal} {Phys. Rev. D}\ }\textbf {\bibinfo {volume}
  {23}},\ \bibinfo {pages} {165} (\bibinfo {year} {1981})}\BibitemShut
  {NoStop}%
\bibitem [{\citenamefont {Bertolini}\ \emph {et~al.}(2004)\citenamefont
  {Bertolini}, \citenamefont {Frigerio},\ and\ \citenamefont
  {Malinsk{\'y}}}]{Bertolini:2004eq}%
  \BibitemOpen
  \bibfield  {author} {\bibinfo {author} {\bibfnamefont {S.}~\bibnamefont
  {Bertolini}}, \bibinfo {author} {\bibfnamefont {M.}~\bibnamefont {Frigerio}},
  \ and\ \bibinfo {author} {\bibfnamefont {M.}~\bibnamefont {Malinsk{\'y}}},\
  }\href {\doibase 10.1103/PhysRevD.70.095002} {\bibfield  {journal} {\bibinfo
  {journal} {Phys. Rev. D}\ }\textbf {\bibinfo {volume} {70}},\ \bibinfo
  {pages} {095002} (\bibinfo {year} {2004})},\ \Eprint
  {http://arxiv.org/abs/hep-ph/0406117}{arXiv:hep-ph/0406117}\BibitemShut
  {NoStop}%
\bibitem [{\citenamefont {Babu}\ and\ \citenamefont
  {Macesanu}(2005)}]{Babu:2005ia}%
  \BibitemOpen
  \bibfield  {author} {\bibinfo {author} {\bibfnamefont {K.~S.}\ \bibnamefont
  {Babu}}\ and\ \bibinfo {author} {\bibfnamefont {C.}~\bibnamefont
  {Macesanu}},\ }\href {\doibase 10.1103/PhysRevD.72.115003} {\bibfield
  {journal} {\bibinfo  {journal} {Phys. Rev. D}\ }\textbf {\bibinfo {volume}
  {72}},\ \bibinfo {pages} {115003} (\bibinfo {year} {2005})},\ \Eprint
  {http://arxiv.org/abs/hep-ph/0505200}{arXiv:hep-ph/0505200}\BibitemShut
  {NoStop}%
\bibitem [{\citenamefont {Bertolini}\ \emph {et~al.}(2006)\citenamefont
  {Bertolini}, \citenamefont {Schwetz},\ and\ \citenamefont
  {Malinsk{\'y}}}]{Bertolini:2006pe}%
  \BibitemOpen
  \bibfield  {author} {\bibinfo {author} {\bibfnamefont {S.}~\bibnamefont
  {Bertolini}}, \bibinfo {author} {\bibfnamefont {T.}~\bibnamefont {Schwetz}},
  \ and\ \bibinfo {author} {\bibfnamefont {M.}~\bibnamefont {Malinsk{\'y}}},\
  }\href {\doibase 10.1103/PhysRevD.73.115012} {\bibfield  {journal} {\bibinfo
  {journal} {Phys. Rev. D}\ }\textbf {\bibinfo {volume} {73}},\ \bibinfo
  {pages} {115012} (\bibinfo {year} {2006})},\ \Eprint
  {http://arxiv.org/abs/hep-ph/0605006}{arXiv:hep-ph/0605006}\BibitemShut
  {NoStop}%
\bibitem [{\citenamefont {Bajc}\ \emph {et~al.}(2008)\citenamefont {Bajc},
  \citenamefont {Dor\v{s}ner},\ and\ \citenamefont
  {Nemev\v{s}ek}}]{Bajc:2008dc}%
  \BibitemOpen
  \bibfield  {author} {\bibinfo {author} {\bibfnamefont {B.}~\bibnamefont
  {Bajc}}, \bibinfo {author} {\bibfnamefont {I.}~\bibnamefont {Dor\v{s}ner}}, \
  and\ \bibinfo {author} {\bibfnamefont {M.}~\bibnamefont {Nemev\v{s}ek}},\
  }\href {\doibase 10.1088/1126-6708/2008/11/007} {\bibfield  {journal}
  {\bibinfo  {journal} {J. High Energy Phys.}\ }\textbf {\bibinfo {volume}
  {11}},\ \bibinfo {pages} {007} (\bibinfo {year} {2008})},\ \Eprint
  {http://arxiv.org/abs/0809.1069}{arXiv:0809.1069 [hep-ph]}\BibitemShut
  {NoStop}%
\bibitem [{\citenamefont {Altarelli}\ and\ \citenamefont
  {Blankenburg}(2011)}]{Altarelli:2010at}%
  \BibitemOpen
  \bibfield  {author} {\bibinfo {author} {\bibfnamefont {G.}~\bibnamefont
  {Altarelli}}\ and\ \bibinfo {author} {\bibfnamefont {G.}~\bibnamefont
  {Blankenburg}},\ }\href {\doibase 10.1007/JHEP03(2011)133} {\bibfield
  {journal} {\bibinfo  {journal} {J. High Energy Phys.}\ }\textbf {\bibinfo
  {volume} {03}},\ \bibinfo {pages} {133} (\bibinfo {year} {2011})},\ \Eprint
  {http://arxiv.org/abs/1012.2697}{arXiv:1012.2697 [hep-ph]}\BibitemShut
  {NoStop}%
\bibitem [{\citenamefont {Fukuyama}\ \emph {et~al.}(2016)\citenamefont
  {Fukuyama}, \citenamefont {Ichikawa},\ and\ \citenamefont
  {Mimura}}]{Fukuyama:2015kra}%
  \BibitemOpen
  \bibfield  {author} {\bibinfo {author} {\bibfnamefont {T.}~\bibnamefont
  {Fukuyama}}, \bibinfo {author} {\bibfnamefont {K.}~\bibnamefont {Ichikawa}},
  \ and\ \bibinfo {author} {\bibfnamefont {Y.}~\bibnamefont {Mimura}},\ }\href
  {\doibase 10.1103/PhysRevD.94.075018} {\bibfield  {journal} {\bibinfo
  {journal} {Phys. Rev. D}\ }\textbf {\bibinfo {volume} {94}},\ \bibinfo
  {pages} {075018} (\bibinfo {year} {2016})},\ \Eprint
  {http://arxiv.org/abs/1508.07078}{arXiv:1508.07078 [hep-ph]}\BibitemShut
  {NoStop}%
\bibitem [{\citenamefont {Altarelli}\ and\ \citenamefont
  {Meloni}(2013)}]{Altarelli:2013aqa}%
  \BibitemOpen
  \bibfield  {author} {\bibinfo {author} {\bibfnamefont {G.}~\bibnamefont
  {Altarelli}}\ and\ \bibinfo {author} {\bibfnamefont {D.}~\bibnamefont
  {Meloni}},\ }\href {\doibase 10.1007/JHEP08(2013)021} {\bibfield  {journal}
  {\bibinfo  {journal} {J. High Energy Phys.}\ }\textbf {\bibinfo {volume}
  {08}},\ \bibinfo {pages} {021} (\bibinfo {year} {2013})},\ \Eprint
  {http://arxiv.org/abs/1305.1001}{arXiv:1305.1001 [hep-ph]}\BibitemShut
  {NoStop}%
\bibitem [{\citenamefont {Dueck}\ and\ \citenamefont
  {Rodejohann}(2013)}]{Dueck:2013gca}%
  \BibitemOpen
  \bibfield  {author} {\bibinfo {author} {\bibfnamefont {A.}~\bibnamefont
  {Dueck}}\ and\ \bibinfo {author} {\bibfnamefont {W.}~\bibnamefont
  {Rodejohann}},\ }\href {\doibase 10.1007/JHEP09(2013)024} {\bibfield
  {journal} {\bibinfo  {journal} {J. High Energy Phys.}\ }\textbf {\bibinfo
  {volume} {09}},\ \bibinfo {pages} {024} (\bibinfo {year} {2013})},\ \Eprint
  {http://arxiv.org/abs/1306.4468}{arXiv:1306.4468 [hep-ph]}\BibitemShut
  {NoStop}%
\bibitem [{\citenamefont {Meloni}\ \emph {et~al.}(2014)\citenamefont {Meloni},
  \citenamefont {Ohlsson},\ and\ \citenamefont {Riad}}]{Meloni:2014rga}%
  \BibitemOpen
  \bibfield  {author} {\bibinfo {author} {\bibfnamefont {D.}~\bibnamefont
  {Meloni}}, \bibinfo {author} {\bibfnamefont {T.}~\bibnamefont {Ohlsson}}, \
  and\ \bibinfo {author} {\bibfnamefont {S.}~\bibnamefont {Riad}},\ }\href
  {\doibase 10.1007/JHEP12(2014)052} {\bibfield  {journal} {\bibinfo  {journal}
  {J. High Energy Phys.}\ }\textbf {\bibinfo {volume} {12}},\ \bibinfo {pages}
  {052} (\bibinfo {year} {2014})},\ \Eprint
  {http://arxiv.org/abs/1409.3730}{arXiv:1409.3730 [hep-ph]}\BibitemShut
  {NoStop}%
\bibitem [{\citenamefont {Babu}\ and\ \citenamefont
  {Khan}(2015)}]{Babu:2015bna}%
  \BibitemOpen
  \bibfield  {author} {\bibinfo {author} {\bibfnamefont {K.~S.}\ \bibnamefont
  {Babu}}\ and\ \bibinfo {author} {\bibfnamefont {S.}~\bibnamefont {Khan}},\
  }\href {\doibase 10.1103/PhysRevD.92.075018} {\bibfield  {journal} {\bibinfo
  {journal} {Phys. Rev. D}\ }\textbf {\bibinfo {volume} {92}},\ \bibinfo
  {pages} {075018} (\bibinfo {year} {2015})},\ \Eprint
  {http://arxiv.org/abs/1507.06712}{arXiv:1507.06712 [hep-ph]}\BibitemShut
  {NoStop}%
\bibitem [{\citenamefont {Meloni}\ \emph {et~al.}(2017)\citenamefont {Meloni},
  \citenamefont {Ohlsson},\ and\ \citenamefont {Riad}}]{Meloni:2016rnt}%
  \BibitemOpen
  \bibfield  {author} {\bibinfo {author} {\bibfnamefont {D.}~\bibnamefont
  {Meloni}}, \bibinfo {author} {\bibfnamefont {T.}~\bibnamefont {Ohlsson}}, \
  and\ \bibinfo {author} {\bibfnamefont {S.}~\bibnamefont {Riad}},\ }\href
  {\doibase 10.1007/JHEP03(2017)045} {\bibfield  {journal} {\bibinfo  {journal}
  {J. High Energy Phys.}\ }\textbf {\bibinfo {volume} {03}},\ \bibinfo {pages}
  {045} (\bibinfo {year} {2017})},\ \Eprint
  {http://arxiv.org/abs/1612.07973}{arXiv:1612.07973 [hep-ph]}\BibitemShut
  {NoStop}%
\bibitem [{\citenamefont {Ohlsson}\ and\ \citenamefont
  {Pernow}(2018)}]{Ohlsson:2018qpt}%
  \BibitemOpen
  \bibfield  {author} {\bibinfo {author} {\bibfnamefont {T.}~\bibnamefont
  {Ohlsson}}\ and\ \bibinfo {author} {\bibfnamefont {M.}~\bibnamefont
  {Pernow}},\ }\href {\doibase 10.1007/JHEP11(2018)028} {\bibfield  {journal}
  {\bibinfo  {journal} {J. High Energy Phys.}\ }\textbf {\bibinfo {volume}
  {11}},\ \bibinfo {pages} {028} (\bibinfo {year} {2018})},\ \Eprint
  {http://arxiv.org/abs/1804.04560}{arXiv:1804.04560 [hep-ph]}\BibitemShut
  {NoStop}%
\bibitem [{\citenamefont {Boucenna}\ \emph {et~al.}(2019)\citenamefont
  {Boucenna}, \citenamefont {Ohlsson},\ and\ \citenamefont
  {Pernow}}]{Boucenna:2018wjc}%
  \BibitemOpen
  \bibfield  {author} {\bibinfo {author} {\bibfnamefont {S.~M.}\ \bibnamefont
  {Boucenna}}, \bibinfo {author} {\bibfnamefont {T.}~\bibnamefont {Ohlsson}}, \
  and\ \bibinfo {author} {\bibfnamefont {M.}~\bibnamefont {Pernow}},\ }\href
  {\doibase 10.1016/j.physletb.2019.03.045} {\bibfield  {journal} {\bibinfo
  {journal} {Phys. Lett. B}\ }\textbf {\bibinfo {volume} {792}},\ \bibinfo
  {pages} {251} (\bibinfo {year} {2019})},\ \Eprint
  {http://arxiv.org/abs/1812.10548}{arXiv:1812.10548 [hep-ph]}\BibitemShut
  {NoStop}%
\bibitem [{\citenamefont {Joshipura}\ and\ \citenamefont
  {Patel}(2011)}]{Joshipura:2011nn}%
  \BibitemOpen
  \bibfield  {author} {\bibinfo {author} {\bibfnamefont {A.~S.}\ \bibnamefont
  {Joshipura}}\ and\ \bibinfo {author} {\bibfnamefont {K.~M.}\ \bibnamefont
  {Patel}},\ }\href {\doibase 10.1103/PhysRevD.83.095002} {\bibfield  {journal}
  {\bibinfo  {journal} {Phys. Rev. D}\ }\textbf {\bibinfo {volume} {83}},\
  \bibinfo {pages} {095002} (\bibinfo {year} {2011})},\ \Eprint
  {http://arxiv.org/abs/1102.5148}{arXiv:1102.5148 [hep-ph]}\BibitemShut
  {NoStop}%
\bibitem [{\citenamefont {Babu}\ \emph {et~al.}(2017)\citenamefont {Babu},
  \citenamefont {Bajc},\ and\ \citenamefont {Saad}}]{Babu:2016bmy}%
  \BibitemOpen
  \bibfield  {author} {\bibinfo {author} {\bibfnamefont {K.~S.}\ \bibnamefont
  {Babu}}, \bibinfo {author} {\bibfnamefont {B.}~\bibnamefont {Bajc}}, \ and\
  \bibinfo {author} {\bibfnamefont {S.}~\bibnamefont {Saad}},\ }\href {\doibase
  10.1007/JHEP02(2017)136} {\bibfield  {journal} {\bibinfo  {journal} {J. High
  Energy Phys.}\ }\textbf {\bibinfo {volume} {02}},\ \bibinfo {pages} {136}
  (\bibinfo {year} {2017})},\ \Eprint
  {http://arxiv.org/abs/1612.04329}{arXiv:1612.04329 [hep-ph]}\BibitemShut
  {NoStop}%
\bibitem [{\citenamefont {Peccei}\ and\ \citenamefont
  {Quinn}(1977{\natexlab{a}})}]{Peccei:1977hh}%
  \BibitemOpen
  \bibfield  {author} {\bibinfo {author} {\bibfnamefont {R.~D.}\ \bibnamefont
  {Peccei}}\ and\ \bibinfo {author} {\bibfnamefont {H.~R.}\ \bibnamefont
  {Quinn}},\ }\href {\doibase 10.1103/PhysRevLett.38.1440} {\bibfield
  {journal} {\bibinfo  {journal} {Phys. Rev. Lett.}\ }\textbf {\bibinfo
  {volume} {38}},\ \bibinfo {pages} {1440} (\bibinfo {year}
  {1977}{\natexlab{a}})}\BibitemShut {NoStop}%
\bibitem [{\citenamefont {Peccei}\ and\ \citenamefont
  {Quinn}(1977{\natexlab{b}})}]{Peccei:1977ur}%
  \BibitemOpen
  \bibfield  {author} {\bibinfo {author} {\bibfnamefont {R.~D.}\ \bibnamefont
  {Peccei}}\ and\ \bibinfo {author} {\bibfnamefont {H.~R.}\ \bibnamefont
  {Quinn}},\ }\href {\doibase 10.1103/PhysRevD.16.1791} {\bibfield  {journal}
  {\bibinfo  {journal} {Phys. Rev. D}\ }\textbf {\bibinfo {volume} {16}},\
  \bibinfo {pages} {1791} (\bibinfo {year} {1977}{\natexlab{b}})}\BibitemShut
  {NoStop}%
\bibitem [{\citenamefont {Weinberg}(1978)}]{Weinberg:1977ma}%
  \BibitemOpen
  \bibfield  {author} {\bibinfo {author} {\bibfnamefont {S.}~\bibnamefont
  {Weinberg}},\ }\href {\doibase 10.1103/PhysRevLett.40.223} {\bibfield
  {journal} {\bibinfo  {journal} {Phys. Rev. Lett.}\ }\textbf {\bibinfo
  {volume} {40}},\ \bibinfo {pages} {223} (\bibinfo {year} {1978})}\BibitemShut
  {NoStop}%
\bibitem [{\citenamefont {Wilczek}(1978)}]{Wilczek:1977pj}%
  \BibitemOpen
  \bibfield  {author} {\bibinfo {author} {\bibfnamefont {F.}~\bibnamefont
  {Wilczek}},\ }\href {\doibase 10.1103/PhysRevLett.40.279} {\bibfield
  {journal} {\bibinfo  {journal} {Phys. Rev. Lett.}\ }\textbf {\bibinfo
  {volume} {40}},\ \bibinfo {pages} {279} (\bibinfo {year} {1978})}\BibitemShut
  {NoStop}%
\bibitem [{\citenamefont {Bajc}\ \emph {et~al.}(2006)\citenamefont {Bajc},
  \citenamefont {Melfo}, \citenamefont {Senjanovi{\'c}},\ and\ \citenamefont
  {Vissani}}]{Bajc:2005zf}%
  \BibitemOpen
  \bibfield  {author} {\bibinfo {author} {\bibfnamefont {B.}~\bibnamefont
  {Bajc}}, \bibinfo {author} {\bibfnamefont {A.}~\bibnamefont {Melfo}},
  \bibinfo {author} {\bibfnamefont {G.}~\bibnamefont {Senjanovi{\'c}}}, \ and\
  \bibinfo {author} {\bibfnamefont {F.}~\bibnamefont {Vissani}},\ }\href
  {\doibase 10.1103/PhysRevD.73.055001} {\bibfield  {journal} {\bibinfo
  {journal} {Phys. Rev. D}\ }\textbf {\bibinfo {volume} {73}},\ \bibinfo
  {pages} {055001} (\bibinfo {year} {2006})},\ \Eprint
  {http://arxiv.org/abs/hep-ph/0510139}{arXiv:hep-ph/0510139}\BibitemShut
  {NoStop}%
\bibitem [{\citenamefont {Frigerio}\ and\ \citenamefont
  {Hambye}(2010)}]{Frigerio:2009wf}%
  \BibitemOpen
  \bibfield  {author} {\bibinfo {author} {\bibfnamefont {M.}~\bibnamefont
  {Frigerio}}\ and\ \bibinfo {author} {\bibfnamefont {T.}~\bibnamefont
  {Hambye}},\ }\href {\doibase 10.1103/PhysRevD.81.075002} {\bibfield
  {journal} {\bibinfo  {journal} {Phys. Rev. D}\ }\textbf {\bibinfo {volume}
  {81}},\ \bibinfo {pages} {075002} (\bibinfo {year} {2010})},\ \Eprint
  {http://arxiv.org/abs/0912.1545}{arXiv:0912.1545 [hep-ph]}\BibitemShut
  {NoStop}%
\bibitem [{\citenamefont {Parida}\ \emph {et~al.}(2017)\citenamefont {Parida},
  \citenamefont {Nayak}, \citenamefont {Satpathy},\ and\ \citenamefont
  {Awasthi}}]{Parida:2016hln}%
  \BibitemOpen
  \bibfield  {author} {\bibinfo {author} {\bibfnamefont {M.~K.}\ \bibnamefont
  {Parida}}, \bibinfo {author} {\bibfnamefont {B.~P.}\ \bibnamefont {Nayak}},
  \bibinfo {author} {\bibfnamefont {R.}~\bibnamefont {Satpathy}}, \ and\
  \bibinfo {author} {\bibfnamefont {R.~L.}\ \bibnamefont {Awasthi}},\ }\href
  {\doibase 10.1007/JHEP04(2017)075} {\bibfield  {journal} {\bibinfo  {journal}
  {J. High Energy Phys.}\ }\textbf {\bibinfo {volume} {04}},\ \bibinfo {pages}
  {075} (\bibinfo {year} {2017})},\ \Eprint
  {http://arxiv.org/abs/1608.03956}{arXiv:1608.03956 [hep-ph]}\BibitemShut
  {NoStop}%
\bibitem [{\citenamefont {Bajc}\ \emph {et~al.}(2003)\citenamefont {Bajc},
  \citenamefont {Senjanovi{\'c}},\ and\ \citenamefont {Vissani}}]{Bajc:2002iw}%
  \BibitemOpen
  \bibfield  {author} {\bibinfo {author} {\bibfnamefont {B.}~\bibnamefont
  {Bajc}}, \bibinfo {author} {\bibfnamefont {G.}~\bibnamefont
  {Senjanovi{\'c}}}, \ and\ \bibinfo {author} {\bibfnamefont {F.}~\bibnamefont
  {Vissani}},\ }\href {\doibase 10.1103/PhysRevLett.90.051802} {\bibfield
  {journal} {\bibinfo  {journal} {Phys. Rev. Lett.}\ }\textbf {\bibinfo
  {volume} {90}},\ \bibinfo {pages} {051802} (\bibinfo {year} {2003})},\
  \Eprint
  {http://arxiv.org/abs/hep-ph/0210207}{arXiv:hep-ph/0210207}\BibitemShut
  {NoStop}%
\bibitem [{\citenamefont {Bajc}\ \emph {et~al.}(2004)\citenamefont {Bajc},
  \citenamefont {Senjanovi{\'c}},\ and\ \citenamefont {Vissani}}]{Bajc:2004fj}%
  \BibitemOpen
  \bibfield  {author} {\bibinfo {author} {\bibfnamefont {B.}~\bibnamefont
  {Bajc}}, \bibinfo {author} {\bibfnamefont {G.}~\bibnamefont
  {Senjanovi{\'c}}}, \ and\ \bibinfo {author} {\bibfnamefont {F.}~\bibnamefont
  {Vissani}},\ }\href {\doibase 10.1103/PhysRevD.70.093002} {\bibfield
  {journal} {\bibinfo  {journal} {Phys. Rev. D}\ }\textbf {\bibinfo {volume}
  {70}},\ \bibinfo {pages} {093002} (\bibinfo {year} {2004})},\ \Eprint
  {http://arxiv.org/abs/hep-ph/0402140}{arXiv:hep-ph/0402140}\BibitemShut
  {NoStop}%
\bibitem [{\citenamefont {Jones}(1982)}]{Jones:1981we}%
  \BibitemOpen
  \bibfield  {author} {\bibinfo {author} {\bibfnamefont {D.~R.~T.}\
  \bibnamefont {Jones}},\ }\href {\doibase 10.1103/PhysRevD.25.581} {\bibfield
  {journal} {\bibinfo  {journal} {Phys. Rev. D}\ }\textbf {\bibinfo {volume}
  {25}},\ \bibinfo {pages} {581} (\bibinfo {year} {1982})}\BibitemShut
  {NoStop}%
\bibitem [{\citenamefont {Machacek}\ and\ \citenamefont
  {Vaughn}(1983)}]{Machacek:1983tz}%
  \BibitemOpen
  \bibfield  {author} {\bibinfo {author} {\bibfnamefont {M.~E.}\ \bibnamefont
  {Machacek}}\ and\ \bibinfo {author} {\bibfnamefont {M.~T.}\ \bibnamefont
  {Vaughn}},\ }\href {\doibase 10.1016/0550-3213(83)90610-7} {\bibfield
  {journal} {\bibinfo  {journal} {Nucl. Phys. B}\ }\textbf {\bibinfo {volume}
  {222}},\ \bibinfo {pages} {83} (\bibinfo {year} {1983})}\BibitemShut
  {NoStop}%
\bibitem [{\citenamefont {Machacek}\ and\ \citenamefont
  {Vaughn}(1984)}]{Machacek:1983fi}%
  \BibitemOpen
  \bibfield  {author} {\bibinfo {author} {\bibfnamefont {M.~E.}\ \bibnamefont
  {Machacek}}\ and\ \bibinfo {author} {\bibfnamefont {M.~T.}\ \bibnamefont
  {Vaughn}},\ }\href {\doibase 10.1016/0550-3213(84)90533-9} {\bibfield
  {journal} {\bibinfo  {journal} {Nucl. Phys. B}\ }\textbf {\bibinfo {volume}
  {236}},\ \bibinfo {pages} {221} (\bibinfo {year} {1984})}\BibitemShut
  {NoStop}%
\bibitem [{\citenamefont {Machacek}\ and\ \citenamefont
  {Vaughn}(1985)}]{Machacek:1984zw}%
  \BibitemOpen
  \bibfield  {author} {\bibinfo {author} {\bibfnamefont {M.~E.}\ \bibnamefont
  {Machacek}}\ and\ \bibinfo {author} {\bibfnamefont {M.~T.}\ \bibnamefont
  {Vaughn}},\ }\href {\doibase 10.1016/0550-3213(85)90040-9} {\bibfield
  {journal} {\bibinfo  {journal} {Nucl. Phys. B}\ }\textbf {\bibinfo {volume}
  {249}},\ \bibinfo {pages} {70} (\bibinfo {year} {1985})}\BibitemShut
  {NoStop}%
\bibitem [{\citenamefont {Antusch}\ \emph {et~al.}(2002)\citenamefont
  {Antusch}, \citenamefont {Kersten}, \citenamefont {Lindner},\ and\
  \citenamefont {Ratz}}]{Antusch:2002rr}%
  \BibitemOpen
  \bibfield  {author} {\bibinfo {author} {\bibfnamefont {S.}~\bibnamefont
  {Antusch}}, \bibinfo {author} {\bibfnamefont {J.}~\bibnamefont {Kersten}},
  \bibinfo {author} {\bibfnamefont {M.}~\bibnamefont {Lindner}}, \ and\
  \bibinfo {author} {\bibfnamefont {M.}~\bibnamefont {Ratz}},\ }\href {\doibase
  10.1016/S0370-2693(02)01960-3} {\bibfield  {journal} {\bibinfo  {journal}
  {Phys. Lett. B}\ }\textbf {\bibinfo {volume} {538}},\ \bibinfo {pages} {87}
  (\bibinfo {year} {2002})},\ \Eprint
  {http://arxiv.org/abs/hep-ph/0203233}{arXiv:hep-ph/0203233}\BibitemShut
  {NoStop}%
\bibitem [{\citenamefont {Antusch}\ \emph {et~al.}(2005)\citenamefont
  {Antusch}, \citenamefont {Kersten}, \citenamefont {Lindner}, \citenamefont
  {Ratz},\ and\ \citenamefont {Schmidt}}]{Antusch:2005gp}%
  \BibitemOpen
  \bibfield  {author} {\bibinfo {author} {\bibfnamefont {S.}~\bibnamefont
  {Antusch}}, \bibinfo {author} {\bibfnamefont {J.}~\bibnamefont {Kersten}},
  \bibinfo {author} {\bibfnamefont {M.}~\bibnamefont {Lindner}}, \bibinfo
  {author} {\bibfnamefont {M.}~\bibnamefont {Ratz}}, \ and\ \bibinfo {author}
  {\bibfnamefont {M.~A.}\ \bibnamefont {Schmidt}},\ }\href {\doibase
  10.1088/1126-6708/2005/03/024} {\bibfield  {journal} {\bibinfo  {journal} {J.
  High Energy Phys.}\ }\textbf {\bibinfo {volume} {03}},\ \bibinfo {pages}
  {024} (\bibinfo {year} {2005})},\ \Eprint
  {http://arxiv.org/abs/hep-ph/0501272}{arXiv:hep-ph/0501272}\BibitemShut
  {NoStop}%
\bibitem [{\citenamefont {Schmidt}(2007)}]{Schmidt:2007nq}%
  \BibitemOpen
  \bibfield  {author} {\bibinfo {author} {\bibfnamefont {M.~A.}\ \bibnamefont
  {Schmidt}},\ }\href {\doibase 10.1103/PhysRevD.85.099903,
  10.1103/PhysRevD.76.073010} {\bibfield  {journal} {\bibinfo  {journal} {Phys.
  Rev. D}\ }\textbf {\bibinfo {volume} {76}},\ \bibinfo {pages} {073010}
  (\bibinfo {year} {2007})},\ \bibinfo {note} {[Erratum: Phys. Rev. D {\bf 85},
  099903 (2012)]},\ \Eprint {http://arxiv.org/abs/0705.3841}{arXiv:0705.3841
  [hep-ph]}\BibitemShut {NoStop}%
\bibitem [{\citenamefont {Chao}\ and\ \citenamefont
  {Zhang}(2007)}]{Chao:2006ye}%
  \BibitemOpen
  \bibfield  {author} {\bibinfo {author} {\bibfnamefont {W.}~\bibnamefont
  {Chao}}\ and\ \bibinfo {author} {\bibfnamefont {H.}~\bibnamefont {Zhang}},\
  }\href {\doibase 10.1103/PhysRevD.75.033003} {\bibfield  {journal} {\bibinfo
  {journal} {Phys. Rev. D}\ }\textbf {\bibinfo {volume} {75}},\ \bibinfo
  {pages} {033003} (\bibinfo {year} {2007})},\ \Eprint
  {http://arxiv.org/abs/hep-ph/0611323}{arXiv:hep-ph/0611323}\BibitemShut
  {NoStop}%
\bibitem [{\citenamefont {Fileviez~Perez}\ \emph {et~al.}(2008)\citenamefont
  {Fileviez~Perez}, \citenamefont {Han}, \citenamefont {Huang}, \citenamefont
  {Li},\ and\ \citenamefont {Wang}}]{Perez:2008ha}%
  \BibitemOpen
  \bibfield  {author} {\bibinfo {author} {\bibfnamefont {P.}~\bibnamefont
  {Fileviez~Perez}}, \bibinfo {author} {\bibfnamefont {T.}~\bibnamefont {Han}},
  \bibinfo {author} {\bibfnamefont {G.-y.}\ \bibnamefont {Huang}}, \bibinfo
  {author} {\bibfnamefont {T.}~\bibnamefont {Li}}, \ and\ \bibinfo {author}
  {\bibfnamefont {K.}~\bibnamefont {Wang}},\ }\href {\doibase
  10.1103/PhysRevD.78.015018} {\bibfield  {journal} {\bibinfo  {journal} {Phys.
  Rev. D}\ }\textbf {\bibinfo {volume} {78}},\ \bibinfo {pages} {015018}
  (\bibinfo {year} {2008})},\ \Eprint
  {http://arxiv.org/abs/0805.3536}{arXiv:0805.3536 [hep-ph]}\BibitemShut
  {NoStop}%
\bibitem [{\citenamefont {Ferreira}\ \emph {et~al.}(2019)\citenamefont
  {Ferreira}, \citenamefont {de~Melo}, \citenamefont {Kovalenko}, \citenamefont
  {Pinheiro},\ and\ \citenamefont {Queiroz}}]{Ferreira:2019qpf}%
  \BibitemOpen
  \bibfield  {author} {\bibinfo {author} {\bibfnamefont {M.~M.}\ \bibnamefont
  {Ferreira}}, \bibinfo {author} {\bibfnamefont {T.~B.}\ \bibnamefont
  {de~Melo}}, \bibinfo {author} {\bibfnamefont {S.}~\bibnamefont {Kovalenko}},
  \bibinfo {author} {\bibfnamefont {P.~R.~D.}\ \bibnamefont {Pinheiro}}, \ and\
  \bibinfo {author} {\bibfnamefont {F.~S.}\ \bibnamefont {Queiroz}},\
  }\href@noop {} {\  (\bibinfo {year} {2019})},\ \Eprint
  {http://arxiv.org/abs/1903.07634}{arXiv:1903.07634 [hep-ph]}\BibitemShut
  {NoStop}%
\bibitem [{\citenamefont {Deppisch}\ \emph {et~al.}(2019)\citenamefont
  {Deppisch}, \citenamefont {Schacht},\ and\ \citenamefont
  {Spinrath}}]{Deppisch:2018flu}%
  \BibitemOpen
  \bibfield  {author} {\bibinfo {author} {\bibfnamefont {T.}~\bibnamefont
  {Deppisch}}, \bibinfo {author} {\bibfnamefont {S.}~\bibnamefont {Schacht}}, \
  and\ \bibinfo {author} {\bibfnamefont {M.}~\bibnamefont {Spinrath}},\ }\href
  {\doibase 10.1007/JHEP01(2019)005} {\bibfield  {journal} {\bibinfo  {journal}
  {J. High Energy Phys.}\ }\textbf {\bibinfo {volume} {01}},\ \bibinfo {pages}
  {005} (\bibinfo {year} {2019})},\ \Eprint
  {http://arxiv.org/abs/1811.02895}{arXiv:1811.02895 [hep-ph]}\BibitemShut
  {NoStop}%
\bibitem [{\citenamefont {de~Salas}\ \emph {et~al.}(2018)\citenamefont
  {de~Salas}, \citenamefont {Forero}, \citenamefont {Ternes}, \citenamefont
  {T{\'o}rtola},\ and\ \citenamefont {Valle}}]{deSalas:2017kay}%
  \BibitemOpen
  \bibfield  {author} {\bibinfo {author} {\bibfnamefont {P.~F.}\ \bibnamefont
  {de~Salas}}, \bibinfo {author} {\bibfnamefont {D.~V.}\ \bibnamefont
  {Forero}}, \bibinfo {author} {\bibfnamefont {C.~A.}\ \bibnamefont {Ternes}},
  \bibinfo {author} {\bibfnamefont {M.}~\bibnamefont {T{\'o}rtola}}, \ and\
  \bibinfo {author} {\bibfnamefont {J.~W.~F.}\ \bibnamefont {Valle}},\ }\href
  {\doibase 10.1016/j.physletb.2018.06.019} {\bibfield  {journal} {\bibinfo
  {journal} {Phys. Lett. B}\ }\textbf {\bibinfo {volume} {782}},\ \bibinfo
  {pages} {633} (\bibinfo {year} {2018})},\ \Eprint
  {http://arxiv.org/abs/1708.01186}{arXiv:1708.01186 [hep-ph]}\BibitemShut
  {NoStop}%
\bibitem [{\citenamefont {Charles}\ \emph {et~al.}(2005)\citenamefont
  {Charles}, \citenamefont {Hocker}, \citenamefont {Lacker}, \citenamefont
  {Laplace}, \citenamefont {Le~Diberder}, \citenamefont {Malcles},
  \citenamefont {Ocariz}, \citenamefont {Pivk},\ and\ \citenamefont
  {Roos}}]{Charles:2004jd}%
  \BibitemOpen
  \bibfield  {author} {\bibinfo {author} {\bibfnamefont {J.}~\bibnamefont
  {Charles}}, \bibinfo {author} {\bibfnamefont {A.}~\bibnamefont {Hocker}},
  \bibinfo {author} {\bibfnamefont {H.}~\bibnamefont {Lacker}}, \bibinfo
  {author} {\bibfnamefont {S.}~\bibnamefont {Laplace}}, \bibinfo {author}
  {\bibfnamefont {F.~R.}\ \bibnamefont {Le~Diberder}}, \bibinfo {author}
  {\bibfnamefont {J.}~\bibnamefont {Malcles}}, \bibinfo {author} {\bibfnamefont
  {J.}~\bibnamefont {Ocariz}}, \bibinfo {author} {\bibfnamefont
  {M.}~\bibnamefont {Pivk}}, \ and\ \bibinfo {author} {\bibfnamefont
  {L.}~\bibnamefont {Roos}} (\bibinfo {collaboration} {CKMfitter Group}),\
  }\href {\doibase 10.1140/epjc/s2005-02169-1} {\bibfield  {journal} {\bibinfo
  {journal} {Eur. Phys. J. C}\ }\textbf {\bibinfo {volume} {41}},\ \bibinfo
  {pages} {1} (\bibinfo {year} {2005})},\ \Eprint
  {http://arxiv.org/abs/hep-ph/0406184}{arXiv:hep-ph/0406184}\BibitemShut
  {NoStop}%
\bibitem [{\citenamefont {Bj{\"o}rkeroth}\ \emph {et~al.}(2017)\citenamefont
  {Bj{\"o}rkeroth}, \citenamefont {de~Anda}, \citenamefont {King},\ and\
  \citenamefont {Perdomo}}]{Bjorkeroth:2017ybg}%
  \BibitemOpen
  \bibfield  {author} {\bibinfo {author} {\bibfnamefont {F.}~\bibnamefont
  {Bj{\"o}rkeroth}}, \bibinfo {author} {\bibfnamefont {F.~J.}\ \bibnamefont
  {de~Anda}}, \bibinfo {author} {\bibfnamefont {S.~F.}\ \bibnamefont {King}}, \
  and\ \bibinfo {author} {\bibfnamefont {E.}~\bibnamefont {Perdomo}},\ }\href
  {\doibase 10.1007/JHEP10(2017)148} {\bibfield  {journal} {\bibinfo  {journal}
  {J. High Energy Phys.}\ }\textbf {\bibinfo {volume} {10}},\ \bibinfo {pages}
  {148} (\bibinfo {year} {2017})},\ \Eprint
  {http://arxiv.org/abs/1705.01555}{arXiv:1705.01555 [hep-ph]}\BibitemShut
  {NoStop}%
\bibitem [{\citenamefont {Martinez}\ \emph {et~al.}(2017)\citenamefont
  {Martinez}, \citenamefont {McKay}, \citenamefont {Farmer}, \citenamefont
  {Scott}, \citenamefont {Roebber}, \citenamefont {Putze},\ and\ \citenamefont
  {Conrad}}]{Workgroup:2017htr}%
  \BibitemOpen
  \bibfield  {author} {\bibinfo {author} {\bibfnamefont {G.~D.}\ \bibnamefont
  {Martinez}}, \bibinfo {author} {\bibfnamefont {J.}~\bibnamefont {McKay}},
  \bibinfo {author} {\bibfnamefont {B.}~\bibnamefont {Farmer}}, \bibinfo
  {author} {\bibfnamefont {P.}~\bibnamefont {Scott}}, \bibinfo {author}
  {\bibfnamefont {E.}~\bibnamefont {Roebber}}, \bibinfo {author} {\bibfnamefont
  {A.}~\bibnamefont {Putze}}, \ and\ \bibinfo {author} {\bibfnamefont
  {J.}~\bibnamefont {Conrad}} (\bibinfo {collaboration} {GAMBIT}),\ }\href
  {\doibase 10.1140/epjc/s10052-017-5274-y} {\bibfield  {journal} {\bibinfo
  {journal} {Eur. Phys. J. C}\ }\textbf {\bibinfo {volume} {77}},\ \bibinfo
  {pages} {761} (\bibinfo {year} {2017})},\ \Eprint
  {http://arxiv.org/abs/1705.07959}{arXiv:1705.07959 [hep-ph]}\BibitemShut
  {NoStop}%
\bibitem [{\citenamefont {Wales}\ and\ \citenamefont {Doye}(1997)}]{bh1997}%
  \BibitemOpen
  \bibfield  {author} {\bibinfo {author} {\bibfnamefont {D.~J.}\ \bibnamefont
  {Wales}}\ and\ \bibinfo {author} {\bibfnamefont {J.~P.~K.}\ \bibnamefont
  {Doye}},\ }\href {\doibase 10.1021/jp970984n} {\bibfield  {journal} {\bibinfo
   {journal} {J. Phys. Chem. A}\ }\textbf {\bibinfo {volume} {101}},\ \bibinfo
  {pages} {5111} (\bibinfo {year} {1997})}\BibitemShut {NoStop}%
\bibitem [{\citenamefont {Jones}\ \emph {et~al.}(01  )\citenamefont {Jones},
  \citenamefont {Oliphant}, \citenamefont {Peterson} \emph
  {et~al.}}]{scipy2001}%
  \BibitemOpen
  \bibfield  {author} {\bibinfo {author} {\bibfnamefont {E.}~\bibnamefont
  {Jones}}, \bibinfo {author} {\bibfnamefont {T.}~\bibnamefont {Oliphant}},
  \bibinfo {author} {\bibfnamefont {P.}~\bibnamefont {Peterson}},  \emph
  {et~al.},\ }\href {http://www.scipy.org/} {\enquote {\bibinfo {title}
  {{SciPy}: Open source scientific tools for {Python}},}\ } (\bibinfo {year}
  {2001--})\BibitemShut {NoStop}%
\bibitem [{\citenamefont {Press}\ \emph {et~al.}(1992)\citenamefont {Press},
  \citenamefont {Teukolsky}, \citenamefont {Vetterling},\ and\ \citenamefont
  {Flannery}}]{Press:1992zz}%
  \BibitemOpen
  \bibfield  {author} {\bibinfo {author} {\bibfnamefont {W.~H.}\ \bibnamefont
  {Press}}, \bibinfo {author} {\bibfnamefont {S.~A.}\ \bibnamefont
  {Teukolsky}}, \bibinfo {author} {\bibfnamefont {W.~T.}\ \bibnamefont
  {Vetterling}}, \ and\ \bibinfo {author} {\bibfnamefont {B.~P.}\ \bibnamefont
  {Flannery}},\ }\href@noop {} {\emph {\bibinfo {title} {{Numerical Recipes in
  C: The Art of Scientific Computing}}}}\ (\bibinfo  {publisher} {Cambridge
  University Press},\ \bibinfo {year} {1992})\BibitemShut {NoStop}%
\bibitem [{\citenamefont {Capozzi}\ \emph {et~al.}(2018)\citenamefont
  {Capozzi}, \citenamefont {Lisi}, \citenamefont {Marrone},\ and\ \citenamefont
  {Palazzo}}]{Capozzi:2018ubv}%
  \BibitemOpen
  \bibfield  {author} {\bibinfo {author} {\bibfnamefont {F.}~\bibnamefont
  {Capozzi}}, \bibinfo {author} {\bibfnamefont {E.}~\bibnamefont {Lisi}},
  \bibinfo {author} {\bibfnamefont {A.}~\bibnamefont {Marrone}}, \ and\
  \bibinfo {author} {\bibfnamefont {A.}~\bibnamefont {Palazzo}},\ }\href
  {\doibase 10.1016/j.ppnp.2018.05.005} {\bibfield  {journal} {\bibinfo
  {journal} {Prog. Part. Nucl. Phys.}\ }\textbf {\bibinfo {volume} {102}},\
  \bibinfo {pages} {48} (\bibinfo {year} {2018})},\ \Eprint
  {http://arxiv.org/abs/1804.09678}{arXiv:1804.09678 [hep-ph]}\BibitemShut
  {NoStop}%
\bibitem [{\citenamefont {Esteban}\ \emph {et~al.}(2019)\citenamefont
  {Esteban}, \citenamefont {Gonzalez-Garcia}, \citenamefont
  {Hernandez-Cabezudo}, \citenamefont {Maltoni},\ and\ \citenamefont
  {Schwetz}}]{Esteban:2018azc}%
  \BibitemOpen
  \bibfield  {author} {\bibinfo {author} {\bibfnamefont {I.}~\bibnamefont
  {Esteban}}, \bibinfo {author} {\bibfnamefont {M.~C.}\ \bibnamefont
  {Gonzalez-Garcia}}, \bibinfo {author} {\bibfnamefont {A.}~\bibnamefont
  {Hernandez-Cabezudo}}, \bibinfo {author} {\bibfnamefont {M.}~\bibnamefont
  {Maltoni}}, \ and\ \bibinfo {author} {\bibfnamefont {T.}~\bibnamefont
  {Schwetz}},\ }\href {\doibase 10.1007/JHEP01(2019)106} {\bibfield  {journal}
  {\bibinfo  {journal} {J. High Energy Phys.}\ }\textbf {\bibinfo {volume}
  {01}},\ \bibinfo {pages} {106} (\bibinfo {year} {2019})},\ \Eprint
  {http://arxiv.org/abs/1811.05487}{arXiv:1811.05487 [hep-ph]}\BibitemShut
  {NoStop}%
\bibitem [{\citenamefont {{NuFIT4.0}}(2018)}]{nu-fit18}%
  \BibitemOpen
  \bibfield  {author} {\bibinfo {author} {\bibnamefont {{NuFIT4.0}}},\ }\href
  {\doibase www.nu-fit.org} {}\bibinfo {howpublished}
  {\url{www.nu-fit.org/?q=node/177}} (\bibinfo {year} {2018})\BibitemShut
  {NoStop}%
\bibitem [{\citenamefont {Tanabashi}\ \emph {et~al.}(2018)\citenamefont
  {Tanabashi} \emph {et~al.}}]{Tanabashi:2018oca}%
  \BibitemOpen
  \bibfield  {author} {\bibinfo {author} {\bibfnamefont {M.}~\bibnamefont
  {Tanabashi}} \emph {et~al.} (\bibinfo {collaboration} {Particle Data
  Group}),\ }\href {\doibase 10.1103/PhysRevD.98.030001} {\bibfield  {journal}
  {\bibinfo  {journal} {Phys. Rev. D}\ }\textbf {\bibinfo {volume} {98}},\
  \bibinfo {pages} {030001} (\bibinfo {year} {2018})}\BibitemShut {NoStop}%
\bibitem [{\citenamefont {Deppisch}\ \emph {et~al.}(2015)\citenamefont
  {Deppisch}, \citenamefont {Gonzalo}, \citenamefont {Patra}, \citenamefont
  {Sahu},\ and\ \citenamefont {Sarkar}}]{Deppisch:2014zta}%
  \BibitemOpen
  \bibfield  {author} {\bibinfo {author} {\bibfnamefont {F.~F.}\ \bibnamefont
  {Deppisch}}, \bibinfo {author} {\bibfnamefont {T.~E.}\ \bibnamefont
  {Gonzalo}}, \bibinfo {author} {\bibfnamefont {S.}~\bibnamefont {Patra}},
  \bibinfo {author} {\bibfnamefont {N.}~\bibnamefont {Sahu}}, \ and\ \bibinfo
  {author} {\bibfnamefont {U.}~\bibnamefont {Sarkar}},\ }\href {\doibase
  10.1103/PhysRevD.91.015018} {\bibfield  {journal} {\bibinfo  {journal} {Phys.
  Rev. D}\ }\textbf {\bibinfo {volume} {91}},\ \bibinfo {pages} {015018}
  (\bibinfo {year} {2015})},\ \Eprint
  {http://arxiv.org/abs/1410.6427}{arXiv:1410.6427 [hep-ph]}\BibitemShut
  {NoStop}%
\bibitem [{\citenamefont {P{\"a}s}\ and\ \citenamefont
  {Rodejohann}(2015)}]{Pas:2015eia}%
  \BibitemOpen
  \bibfield  {author} {\bibinfo {author} {\bibfnamefont {H.}~\bibnamefont
  {P{\"a}s}}\ and\ \bibinfo {author} {\bibfnamefont {W.}~\bibnamefont
  {Rodejohann}},\ }\href {\doibase 10.1088/1367-2630/17/11/115010} {\bibfield
  {journal} {\bibinfo  {journal} {New J. Phys.}\ }\textbf {\bibinfo {volume}
  {17}},\ \bibinfo {pages} {115010} (\bibinfo {year} {2015})},\ \Eprint
  {http://arxiv.org/abs/1507.00170}{arXiv:1507.00170 [hep-ph]}\BibitemShut
  {NoStop}%
\end{thebibliography}%

\end{document}